\documentclass[conf]{new-aiaa}

\usepackage[utf8]{inputenc}

\usepackage{graphicx}
\usepackage{amsmath}
\usepackage[version=4]{mhchem}
\usepackage{siunitx}
\usepackage{longtable,tabularx}

\usepackage{mathtools}
\usepackage{comment}
\usepackage{subcaption}
\usepackage{placeins}
\usepackage{tcolorbox}
\title{Overcoming Fundamental Control Limitations in Powered Descent Flight through Robust Incremental Nonlinear Dynamic Inversion}

\author{João T. Diz\footnote{PhD Candidate, Control \& Simulation Section/ GNC Engineer, GNC Team, AIAA Student Member; jtmarquesdossa@tudelft.nl (Corresponding Author).}}
\affil{Delft University of Technology and INDRA Space}
\author{Paulo Rosa\footnote{Head of Flight Segment Engineering.}}
\affil{INDRA Space, 1070-061 Lisbon, Portugal}
\author{Pedro Simplício\footnote{GNC Systems Engineer, Guidance, Navigation and Control Systems Architecture Section (TEC-ECG).} and Paul Acquatella\footnote{GNC Systems Engineer, Guidance, Navigation and Control Systems Architecture Section (TEC-ECG), AIAA Member.}}
\affil{European Space Agency, 2201 AZ Noordwijk, The Netherlands}
\author{Erik-Jan van Kampen\footnote{Associate Professor, Control \& Simulation Section, Faculty of Aerospace Engineering.} and Spilios Theodoulis\footnote{Associate Professor, Control \& Simulation Section, Faculty of Aerospace Engineering, Associate Fellow AIAA Member.}}
\affil{Delft University of Technology, 2628 CD Delft, The Netherlands}
\begin{document}

\maketitle
{\renewcommand{\thefootnote}{}\footnotetext{Accepted for presentation at the AIAA SciTech 2027 Forum, Orlando, FL, 11--15 January 2027, and to be published by the American Institute of Aeronautics and Astronautics, Inc. Copyright \copyright\ 2026 by the authors.}}

\begin{abstract}
\boldmath
This work identifies a fundamental, controller-independent limitation in reusable launch vehicle powered descent:  a sudden $180^\circ$ phase transition in the open-loop transfer functions from the control inputs to the lateral drift and drift rate. The transition occurs inside the flight envelope, reverses the effective control direction, and can destabilize the closed loop system under uncertainty. An analytical condition governing this transition is derived, its physical mechanism is established, and its dependence on vehicle mass, aerodynamic properties, and geometry is characterized. An Incremental Nonlinear Dynamic Inversion (INDI)-based control architecture is then developed to inherently address this limitation and is tuned via $\mathcal{H}_\infty$ Closed Loop Shaping to provide local robustness and performance guarantees. The internal dynamics introduced by the INDI inversion are analytically derived and proven to be stable throughout the flight envelope. The proposed controller is compared against a robust gain-scheduled baseline controller through linear stability analysis and an extensive nonlinear Monte Carlo campaign. Both controllers achieve $100\%$ success across the considered Monte Carlo campaign, though the INDI-based controller provides improved tracking performance, superior nominal and dispersed stability margins, requires fewer controller design points, and reduces controller tuning complexity. 
\end{abstract}

\section{Introduction}
\lettrine{P}{owered} descent and landing of reusable launch vehicles (RLVs) have become a major focus of research, driven by the growing demand for rapid, cost-effective, and reliable access to space \cite{INDI_Simplicio2025}. The demonstrated benefits of recovering and reusing launch vehicles have further motivated the development of guidance, navigation, and control (GNC) technologies capable of achieving precise and robust autonomous landing.

\par From a control standpoint, RLV descent and landing presents a particularly demanding problem. The vehicle exhibits highly nonlinear and strongly time-varying dynamics, resulting from effects such as plume-flow interactions and substantial changes in mass, center of gravity, and inertia (MCI) throughout descent. Multiple actuation systems are available, including thrust vector control (TVC), aerodynamic surfaces (fins), and reaction control systems (RCS), but their control authority is both limited and dependent on the flight condition. In particular, the effectiveness of the aerodynamic surfaces decreases progressively as the vehicle decelerates toward landing.

\par The difficulties associated with the nonlinear and time-varying dynamics are further exacerbated by significant uncertainties, including aerodynamic coefficients, thrust magnitude and misalignments, atmospheric disturbances, and MCI properties. At the same time, the controller must satisfy stringent terminal constraints at touchdown, including requirements on landing position and velocity, attitude, and angular rates. The problem is further complicated by online trajectory generation and guidance updates, which require the control system to accurately track a broad range of trajectories while maintaining performance under dispersed operating conditions.

\par Several control methodologies have been investigated for RLV descent and landing, as well as for the closely related atmospheric re-entry problem. Linear Quadratic Regulator (LQR) approaches have been extensively employed for winged reusable and re-entry vehicles \cite{Roenneke1992,Roenneke1993,Mooij1998}, while Nonlinear Dynamic Inversion (NDI) \cite{roenneke1995nonlinear,recasens2005robust,juliana2004flight}, Sliding Mode Control (SMC) \cite{hall2006sliding,shtessel2000reusable,shtessel2002reusable}, and Model Predictive Control (MPC) \cite{zaragoza2025decreasing,Pascucci2015} have also been investigated for re-entry and landing applications. In the context of launch vehicle control, $\mathcal{H}_{\infty}$-based methods have been applied to ascent control \cite{Navarro2019,simplicio2019,rocond,diz2026thrust} and subsequently to descent and powered landing \cite{sagliano2021robust,sagliano2022structured,guadagnini2025end}. These approaches provide local robustness and performance guarantees (\textit{i.e.}, around the operating conditions of the linear time-invariant models used for synthesis), reducing reliance on extensive post-design Verification and Validation (V\&V). More recently, inversion-based architectures, including NDI and Incremental NDI (INDI), have been combined with structured $\mathcal{H}_{\infty}$ synthesis to exploit the advantages of both approaches \cite{pollack,encarnaccao2025tuning,encarnaccao2026multi,Diz_2026_esagnc,Diz_2026_eurognc}. The inversion-based structure can reduce the need for extensive gain-scheduling, while structured $\mathcal{H}_{\infty}$ synthesis provides a systematic framework for tuning the resulting architecture and establishing local robustness and performance guarantees.

\par Our most recent work in this line of research applied the methodology to the lateral position and attitude control of the RETALT reusable launch vehicle during powered descent and landing \cite{Diz_2026_esagnc}. An INDI-based control architecture was tuned using structured $\mathcal{H}_{\infty}$ synthesis and achieved a near-$100\%$ success rate in an initial Monte Carlo campaign, while reducing the number of gain-scheduling flight points from more than twenty in the original control system down to three. However, extended robustness analyses under larger uncertainty dispersions revealed a previously unobserved limitation associated with the vehicle dynamics: the open-loop transfer functions from the control inputs to drift and drift rate undergo an abrupt $180^\circ$ phase transition within the flight envelope, requiring the corresponding controller gains to change sign in order to maintain stability. Because the location of this transition is sensitive to aerodynamic and mass uncertainties, the scheduled gain sign changes may occur prematurely or too late relative to the actual vehicle dynamics, resulting in transient instability and degraded closed loop performance. This limitation, which to the authors' knowledge has not been reported in the open literature, motivates the work presented in this paper.

\par This work addresses the aforementioned limitation by investigating its underlying mechanism and developing an INDI-based architecture that inherently addresses the resulting control challenge. \textit{The contributions of this paper are threefold:}
\begin{itemize}
    \item An analytical condition governing the occurrence of the $180^\circ$ phase transition is derived, together with a physical interpretation of its origin and an analysis of its dependence on vehicle mass, aerodynamic properties, and geometry;
    \item A new INDI-based architecture is developed and analytically shown to inherently address the phase transition, and the internal dynamics introduced by its feedback linearization are derived and proven to be stable throughout the flight envelope;
    \item The proposed architecture is evaluated through linear and nonlinear analyses and compared against a robust gain-scheduled baseline controller, demonstrating improved trajectory tracking, stability, and robustness while reducing controller design complexity and attaining a $100\%$ success rate in an extensive Monte Carlo campaign.
\end{itemize}

The remainder of the paper is structured as follows: Section~\ref{sec:model} introduces the reusable launch vehicle model employed throughout the study. Section~\ref{sec:l1} derives the analytical condition governing the phase transition and examines its physical origin and dependence on vehicle mass, aerodynamic properties, and geometry. Section~\ref{sec:indi} introduces and tunes the proposed INDI-based architecture, provides the analytical explanation for its inherent handling of the phase transition, and derives and assesses its internal dynamics. Section~\ref{sec:Comparison} introduces a baseline robust controller based on the architecture proposed in \cite{guadagnini2025end} and evaluates both controllers through linear and nonlinear analyses. Finally, Section~\ref{sec:conclusions} summarizes the main conclusions of this work.

\section{Reusable Launch Vehicle Flight Dynamics}
\label{sec:model}
In this section, the reusable launch vehicle model is presented. First, in Subsection~\ref{sec:nonlinear_model}, the full nonlinear rigid-body equations of motion are introduced. Subsequently, in Subsection~\ref{sec:linear_model}, these equations are linearized about the reference trajectory, in a manner suitable for control design and analysis. In addition, the auxiliary components used for linear analysis and controller synthesis are also introduced.
\subsection{Nonlinear Model}
\label{sec:nonlinear_model}
The linear acceleration $\ddot{\mathbf{r}}_{\mathrm{I}}(t)$ of a reusable launch vehicle in an inertial frame (assuming a flat, non-rotating Earth) is given by \cite{simplicio2019}:
\begin{equation}
    \ddot{\mathbf{r}}_{\mathrm{I}}(t)=\dot{\mathbf{v}}_{\mathrm{I}}(t)=\mathbf{g}_{\mathrm{I}}(t)+\frac{1}{m(t)}\left[\mathbf{F}_{\text {aero}, \mathrm{I}}(t)+\mathbf{F}_{\mathrm{TVC}, \mathrm{I}}(t)+\mathbf{F}_{\text {FIN}, \mathrm{I}}(t)+\mathbf{F}_{\text {thr}, \mathrm{I}}(t)\right]
\end{equation}
where $\mathbf{r}_{\mathrm{I}}(t)$ and $\mathbf{v}_{\mathrm{I}}(t)$ denote the vehicle position and velocity with respect to the inertial frame, respectively, and are expressed in inertial coordinates. The terms $\mathbf{F}_{\mathrm{aero,I}}(t)$, $\mathbf{F}_{\mathrm{TVC,I}}(t)$, $\mathbf{F}_{\mathrm{FIN,I}}(t)$, and $\mathbf{F}_{\mathrm{thr,I}}(t)$ represent aerodynamic, thrust-vectoring, fin-induced, and reaction-control-system-induced forces. The time-varying vehicle mass is denoted by $m(t)$, while $\mathbf{g}_{\mathrm{I}}(t)$ accounts for gravitational acceleration in the inertial frame.
\par Similarly, the angular acceleration of the vehicle expressed in the body-fixed frame is given by \cite{simplicio2019}:
\begin{equation}
\begin{aligned}
\dot{\boldsymbol{\omega}}_{\mathrm{B}}(t)=\boldsymbol{J}^{-1}(t)\left[\mathbf{M}_{\text {aero, } \mathrm{B}}(t)+\mathbf{M}_{\mathrm{TVC}, \mathrm{~B}}(t)+\mathbf{M}_{\mathrm{FIN}, \mathrm{~B}}(t)\right. & +\mathbf{M}_{\mathrm{thr}, \mathrm{~B}}(t) \\
& \left.-\boldsymbol{\omega}_{\mathrm{B}}(t) \times \boldsymbol{J}(t) \boldsymbol{\omega}_{\mathrm{B}}(t)-\dot{\boldsymbol{J}}(t) \boldsymbol{\omega}_{\mathrm{B}}(t)\right]
\end{aligned}
\end{equation}
where $\boldsymbol{\omega}_{\mathrm{B}}(t)$ denotes the body angular velocity, $\boldsymbol{J}(t)$ is the inertia tensor, and $\dot{\boldsymbol{J}}(t)$ its time derivative. The moment contributions $\mathbf{M}_{\mathrm{aero,B}}(t)$, $\mathbf{M}_{\mathrm{TVC,B}}(t)$, $\mathbf{M}_{\mathrm{FIN,B}}(t)$, and $\mathbf{M}_{\mathrm{thr,B}}(t)$ correspond respectively to aerodynamic, thrust-vectoring, fin-induced, and reaction-control-system-induced moments, all expressed in the body frame. 
\begin{figure}[b!]
    \centering
    \includegraphics[scale=0.125]{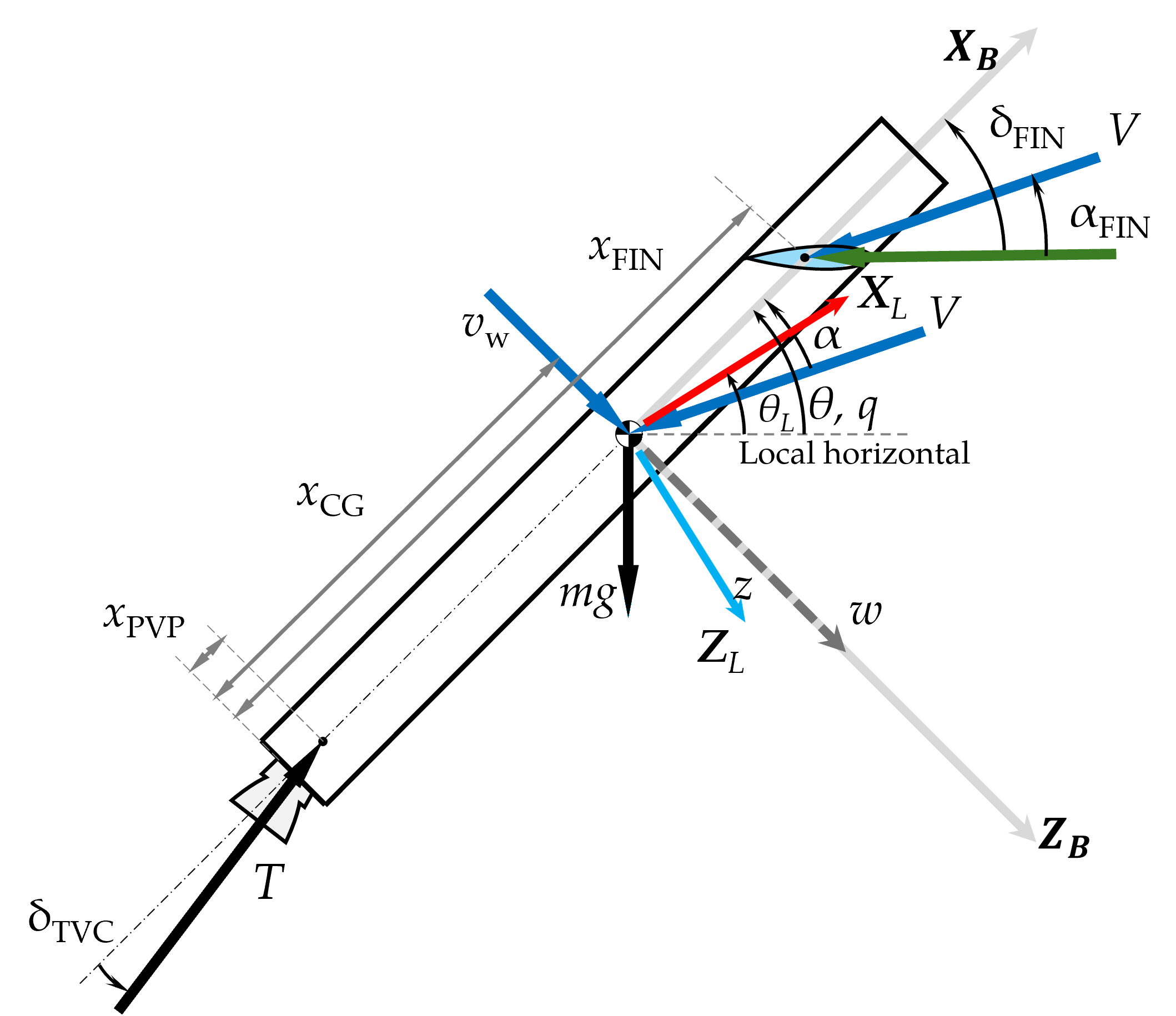}
    \caption{Pitch plane diagram of the reusable launch vehicle.}
    \label{fig:LVPitchDiagram}
\end{figure}
\par For axisymmetric vehicles with quasi-zero roll rate, pitch and yaw dynamics can be assumed as being uncoupled and the task of control design can be performed for each axis separately \cite{roux2007}. The relevant quantities associated with the pitch axis of the reusable launch vehicle are illustrated in Fig.~\ref{fig:LVPitchDiagram}. Under these assumptions, and neglecting the reaction control system, as it is typically not used for lateral position or pitch/yaw control in this phase of flight, the translational and attitude dynamics equations of motion can be expressed in the body-fixed frame as~\cite{simplicio2019}: 
\begin{equation}
\label{EOM}
\begin{aligned}
    m(\dot{w}- q\dot{x}) & = -N_\alpha \alpha - N_f\,\alpha_{\mathrm{FIN}} - T \sin \delta_{\mathrm{TVC}} + m g \cos\theta \\
    J_N \ddot{\theta}&= l_\alpha N_\alpha \alpha + l_f N_f\,\alpha_{\mathrm{FIN}} - l_{{c}} T \sin \delta_{\mathrm{TVC}},
    \qquad \alpha_{\mathrm{FIN}} = \alpha + \delta_{\mathrm{FIN}}
\end{aligned}
\end{equation}
where
\begin{equation*}
\begin{gathered}
l_\alpha = x_{\mathrm{CP}}-x_{\mathrm{CG}}, \qquad
l_{{f}} = x_{\mathrm{FIN}}-x_{\mathrm{CG}}, \qquad
l_{{c}} = x_{\mathrm{CG}}-x_{\mathrm{PVP}}, \\[3pt]
N_\alpha = Q S_{\mathrm{ref}} C_{N_\alpha}, \qquad
N_f = 2 Q S_{\mathrm{FIN}} C_{f_\alpha},
\end{gathered}
\end{equation*}
and $w$ denotes the lateral velocity in the body frame, $\dot{x}$ the longitudinal velocity, and $q = \dot{\theta}$ the pitch rate. The dynamic pressure is denoted by $Q$, while $S_{\mathrm{ref}}$ and $S_{\mathrm{FIN}}$ denote the vehicle reference area and fin reference area, respectively. The coefficient $C_{N_\alpha}$ represents the vehicle normal force gradient, and $C_{f_{\alpha}}$ the fin normal force gradient. The angles $\alpha$ and $\alpha_{\mathrm{FIN}}$ denote the angle of attack (AoA) of the RLV and of the fins, respectively, while $\delta_{\mathrm{FIN}}$ and $\delta_{\mathrm{TVC}}$ are the fin and engine deflections. The fin angle of attack is given by $\alpha_{\mathrm{FIN}}=\alpha+\delta_{\mathrm{FIN}}$; note that this fin deflection convention is opposite to that of \cite{simplicio2019}, \textit{i.e.} $\delta_{\mathrm{FIN}}=-\beta_{\mathrm{fin}}$. The angle of attack depends on the wind speed, shown in Fig.~\ref{fig:LVPitchDiagram} as $v_w$. The parameters $m$, $g$, $\theta$, and $J_N$ denote the vehicle mass, gravitational acceleration, pitch angle with respect to the local horizontal, and normal moment of inertia. The lever arms $l_\alpha$, $l_f$, and $l_c$ are defined with respect to the positions of the center of pressure ($x_{\mathrm{CP}}$), center of gravity ($x_{\mathrm{CG}}$), fin location ($x_{\mathrm{FIN}}$), and nozzle pivot point ($x_{\mathrm{PVP}}$). 
To account for the lateral motion of the vehicle relative to the reference trajectory, as is customary in launch vehicle control \cite{Navarro2019,JGCD_Diz}, the model is augmented with drift and drift rate states, describing the lateral position and velocity of the vehicle in the trajectory frame ($x_L$ and $z_L$ in Fig.~\ref{fig:LVPitchDiagram}). The $x_L$ axis of this frame is aligned with the reference trajectory and inclined by $\theta_L$ with respect to the local horizontal. Since this frame rotates slowly compared with the vehicle dynamics along the reference trajectory, its rotation is neglected ($\dot{\theta}_L \approx 0$). Following Chapter~5 of \cite{Diz2025}, and since the vehicle remains approximately aligned with the trajectory frame, such that small-angle approximations can be applied to $\theta-\theta_L$, the drift kinematics are given by:
\begin{equation*}
\dot{z} = w - (\theta-\theta_L)\dot{x} + \phi v, \qquad \ddot{z} = \dot{w} - \dot{\theta}\dot{x} - (\theta-\theta_L)\ddot{x} + \dot{\phi}v + \phi\dot{v}
\end{equation*}
where $\phi$ is the roll angle of the vehicle and $v$ is the lateral velocity in the yaw plane. Owing to the axisymmetric nature of the vehicle, together with the assumption of negligible roll motion, the terms $\dot{\phi}v$ and $\phi\dot{v}$ are neglected \cite{Diz2025}. Substituting the resulting kinematic relation into Eq.~\eqref{EOM}, the equations of motion can be expressed in the following control-affine form:
\begin{equation}
\begin{aligned}
\begin{bmatrix}
\ddot{z} \\[1ex]
\dot{q}
\end{bmatrix}
=
\underbrace{\begin{bmatrix}
-(\theta-\theta_L)\ddot{x}
-n_{\alpha'}\alpha
+ g\cos\theta \\[1ex]
\mu_{\alpha'} \alpha
\end{bmatrix}}_{f(x)}
+
\underbrace{\begin{bmatrix}
-n_c \\[1ex]
-\mu_c
\end{bmatrix}}_{g_\text{TVC}(x)}
\sin\delta_{\mathrm{TVC}}
+
\underbrace{\begin{bmatrix}
-n_f \\[1ex]
\mu_f
\end{bmatrix}}_{g_\text{FIN}(x)}
\delta_{\mathrm{FIN}}
\end{aligned}
\label{CA_final}
\end{equation}
where the shorthands above are defined as:
\begin{equation}
\begin{aligned}
n_{\alpha'} &= \frac{N_{\alpha'}}{m},
&\qquad n_c &= \frac{T}{m},
&\qquad n_f &= \frac{N_f}{m}, \\[3pt]
\mu_\alpha &= \frac{l_\alpha N_\alpha}{J_N},
&\qquad \mu_c &= \frac{l_c T}{J_N},
&\qquad \mu_f &= \frac{l_f N_f}{J_N}, \\[3pt]
N_{\alpha'} &= N_\alpha + N_f,
&\qquad \mu_{\alpha'} &= \mu_\alpha + \mu_f .
\end{aligned}
\label{coeffs2}
\end{equation}

\subsection{Linearized Model}
\label{sec:linear_model}
Following the procedure described in \cite{simplicio2019}, Eq.~\eqref{CA_final} can be rewritten in perturbed form, where each variable is expressed as the sum of a steady-state component and a small perturbation ($\nu = \nu_0+\Delta\nu$). This leads to the following nonlinear perturbed dynamics:
\begin{equation}
\begin{aligned}
m\left(\Delta\ddot{z} + (\theta_0-\theta_L)\ddot{x}_0 + \Delta\theta\,\ddot{x}_0 + (\theta_0-\theta_L)\Delta\ddot{x} + \Delta\theta\,\Delta\ddot{x}\right) &= -N_\alpha\left(\alpha_0 + \Delta\alpha\right) - N_f\left(\alpha_0 + \Delta\alpha + \delta_{\mathrm{FIN},0} + \Delta\delta_{\mathrm{FIN}}\right) \\
&\quad - T\sin\left(\delta_{\mathrm{TVC},0} + \Delta\delta_{\mathrm{TVC}}\right) + mg\cos\left(\theta_0 + \Delta\theta\right) \\
J_N\,\Delta\dot{q} &= l_\alpha N_\alpha\left(\alpha_0 + \Delta\alpha\right) + l_f N_f\left(\alpha_0 + \Delta\alpha + \delta_{\mathrm{FIN},0} + \Delta\delta_{\mathrm{FIN}}\right) \\
&\quad - l_c T\sin\left(\delta_{\mathrm{TVC},0} + \Delta\delta_{\mathrm{TVC}}\right)
\end{aligned}
\end{equation}
These equations can be simplified using standard small-perturbation assumptions and other commonly adopted assumptions in launch vehicle applications. First, since the reference angle of attack is small, the vehicle remains approximately aligned with the trajectory frame along the reference trajectory, $\theta_0\approx\theta_L$, such that the terms $(\theta_0-\theta_L)\ddot{x}_0$ and $(\theta_0-\theta_L)\Delta\ddot{x}$ are negligible. Accordingly, the angle of attack ($\alpha = \arctan{\frac{w-v_w}{\dot{x}}}$) is approximated, for perturbations, as:
\begin{equation}
    \Delta \alpha =  \frac{\Delta w-v_w}{\dot{x}_0} = \frac{\Delta \dot{z}+\Delta\theta\dot{x}_0-v_w}{\dot{x}_0}
\end{equation}
Additional first-order approximations are applied to the trigonometric terms and higher-order perturbation products are neglected. Furthermore, the coupling term $\Delta\theta\,\ddot{x}_0$ is neglected by assuming that the nominal longitudinal
acceleration is approximately zero during powered landing,
\begin{equation}
    \frac{T-D}{m}-g\sin\theta_0 \approx 0,
    \label{eq:forcebalance}
\end{equation}
as the vehicle operates close to a longitudinal force-balanced condition, particularly during the latter part of the landing envelope\footnote{While this assumption does not hold exactly throughout the entire flight envelope, it provides a useful simplification of the model. Numerical verification of the resulting LTI models shows good agreement with the underlying nonlinear dynamics.}. Finally, omitting the perturbation notation $\Delta$ for readability, the resulting linear state-space representation can be written as (see Appendix C of \cite{simplicio2019} for details):
\begin{equation}
\left[
\begin{array}{c}
\dot{\theta}\\
\dot{q}\\
\dot{z}\\
\ddot{z}
\end{array}
\right]
=
\left[
\begin{array}{cccc}
0 & 1 & 0 & 0\\
\mu_{\alpha'} & 0 & 0 & \frac{\mu_{\alpha'}}{\dot{x}_0}\\
0 & 0 & 0 & 1 \\
-\frac{N_{\alpha'}}{m}-a_0 & 0 & 0 & -\frac{N_{\alpha'}}{m \dot{x}_0}
\end{array}
\right]
\left[
\begin{array}{c}
\theta\\
q\\
z\\
\dot{z}
\end{array}
\right]
+
\left[
\begin{array}{cc}
0 & 0\\
-\mu_c & \mu_f\\
0 & 0 \\
-\frac{T}{m} & -\frac{N_f}{m}
\end{array}
\right]
\left[
\begin{array}{c}
\delta_{\mathrm{TVC}}\\
\delta_{\mathrm{FIN}}
\end{array}
\right]
+
\left[
\begin{array}{c}
0\\
-\frac{\mu_{\alpha'}}{\dot{x}_0}\\
0 \\
\frac{N_{\alpha'}}{m \dot{x}_0}
\end{array}
\right]v_w 
\label{linEOM}
\end{equation}
where $a_0$ is defined as:
\begin{equation}
\label{ShortHand}
\begin{aligned}
     a_0 = g\sin(\theta_0)
\end{aligned}
\end{equation}
which is strictly positive provided that $0<\theta_0<\pi$, a condition satisfied throughout powered descent since the vehicle remains close to vertical ($\theta_0 \approx \pi/2$). The resulting linear model is further augmented with second-order dynamic models for both the thrust vector control system ($G_\text{TVC}$) and the fin actuators ($G_\text{FIN}$) to capture their bandwidth limitations. In addition, a second-order Padé approximation is included to represent the effects of digital implementation and sampling ($G_\text{Delay-TVC}$ and $G_\text{Delay-FIN}$). The complete augmented system, used for analysis and control design, is given by $G_\text{RLV}$ as illustrated in Fig.~\ref{fig:InversionScheme} (Section~\ref{sec:indi}).

\section{Phase Transition Mechanism}
\label{sec:l1}
Before designing a controller, it is important to first establish the limitations imposed by the plant itself. Fundamental properties such as right-half-plane (RHP) zeros and poles, time delays, and input constraints can restrict the achievable closed loop performance \cite{skogestad2005,seron1997fundamental}. In this section, a limitation arising from the dynamic evolution of the reusable launch vehicle during powered descent is investigated. For our analyses, the reusable launch vehicle parameters from RETALT are used \cite{guadagnini2025end,Diz_2026_esagnc}. 
\par
\begin{figure}[t!]
    \centering

    \begin{subfigure}{0.49\textwidth}
        \centering
        \includegraphics[width=\textwidth]{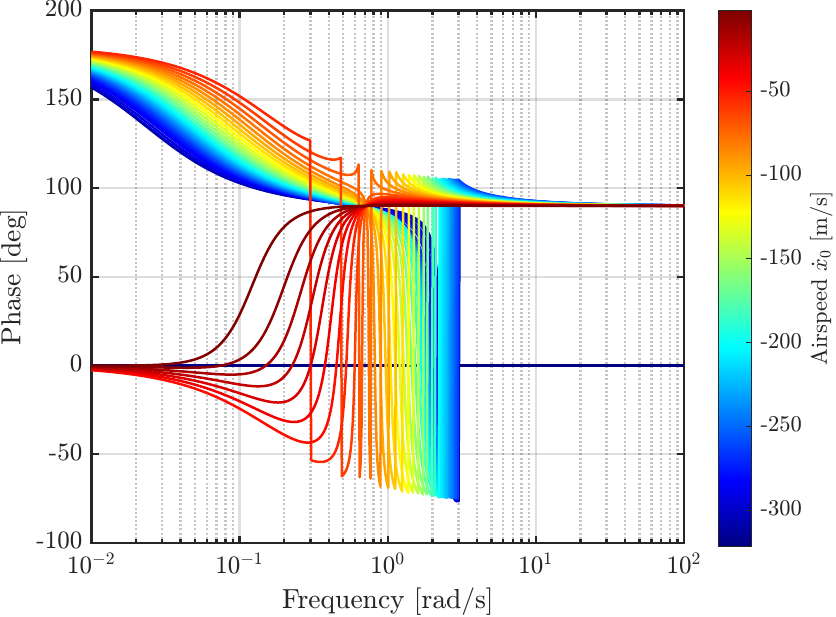}    \caption{$T_{\delta_{\mathrm{TVC}}\rightarrow \dot{z}}$}
    \end{subfigure}
    \hfill
    \begin{subfigure}{0.49\textwidth}
        \centering
        \includegraphics[width=\textwidth]{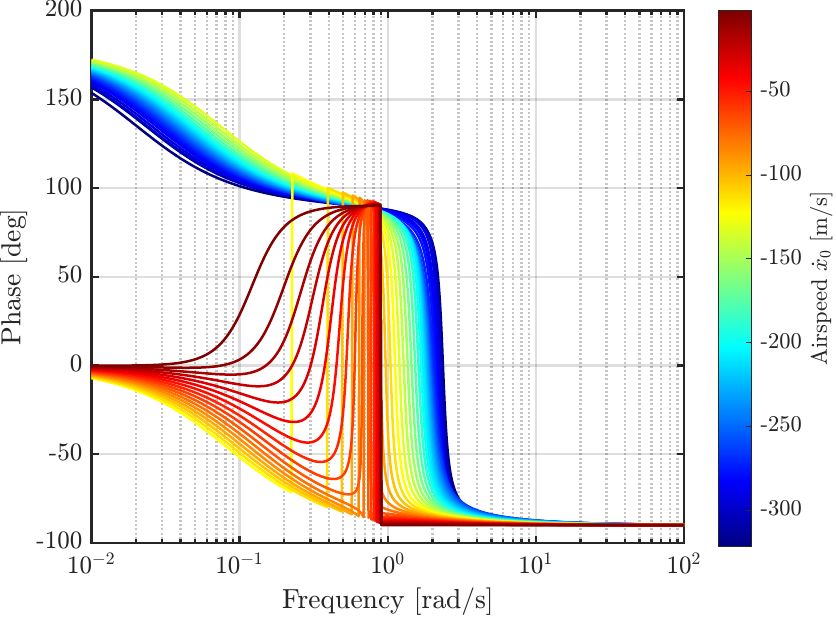}   \caption{$T_{\delta_{\mathrm{FIN}}\rightarrow \dot{z}}$}
    \end{subfigure}

    \caption{Phase of the frequency responses versus airspeed.}
    \label{fig:threeplots}
\end{figure}

 Frequency-domain analyses were conducted by evaluating the evolution of the
input-output Bode plots as a function of airspeed. The most significant observation
emerged from the low-frequency behavior of the system. Figure~\ref{fig:threeplots}
illustrates the phase responses of two open-loop transfer functions, namely
$T_{\delta_{\mathrm{TVC}}\rightarrow\dot{z}}$ and
$T_{\delta_{\mathrm{FIN}}\rightarrow\dot{z}}$, evaluated at different
airspeeds.
\par In both cases, \textit{the phase response exhibits an abrupt transition}
of approximately $180^\circ$ at a critical airspeed. The same phenomenon is
observed in the drift dynamics, whereas no equivalent behavior appears in
the attitude- or attitude-rate-related channels. This phase inversion
necessitated a change in the sign of the feedback gains of the INDI-based
controller developed in \cite{Diz_2026_esagnc} to preserve closed loop
stability and maintain a consistent control action.

Sections~\ref{TVC_PhaseChange} and~\ref{Fins_PhaseChange} derive analytical expressions that identify the conditions under which this transition occurs for the TVC and fin inputs, respectively, and Section~\ref{Alloc_PhaseChange} extends them to a control input distributed to both actuators through a control allocation. Section~\ref{sec:discussion} then discusses the underlying physical mechanism and its implications for control and vehicle design.

\subsection{Underlying Mechanism - TVC}
\label{TVC_PhaseChange}
\begin{figure}[b!]
    \centering
    \includegraphics[width=0.7\linewidth]{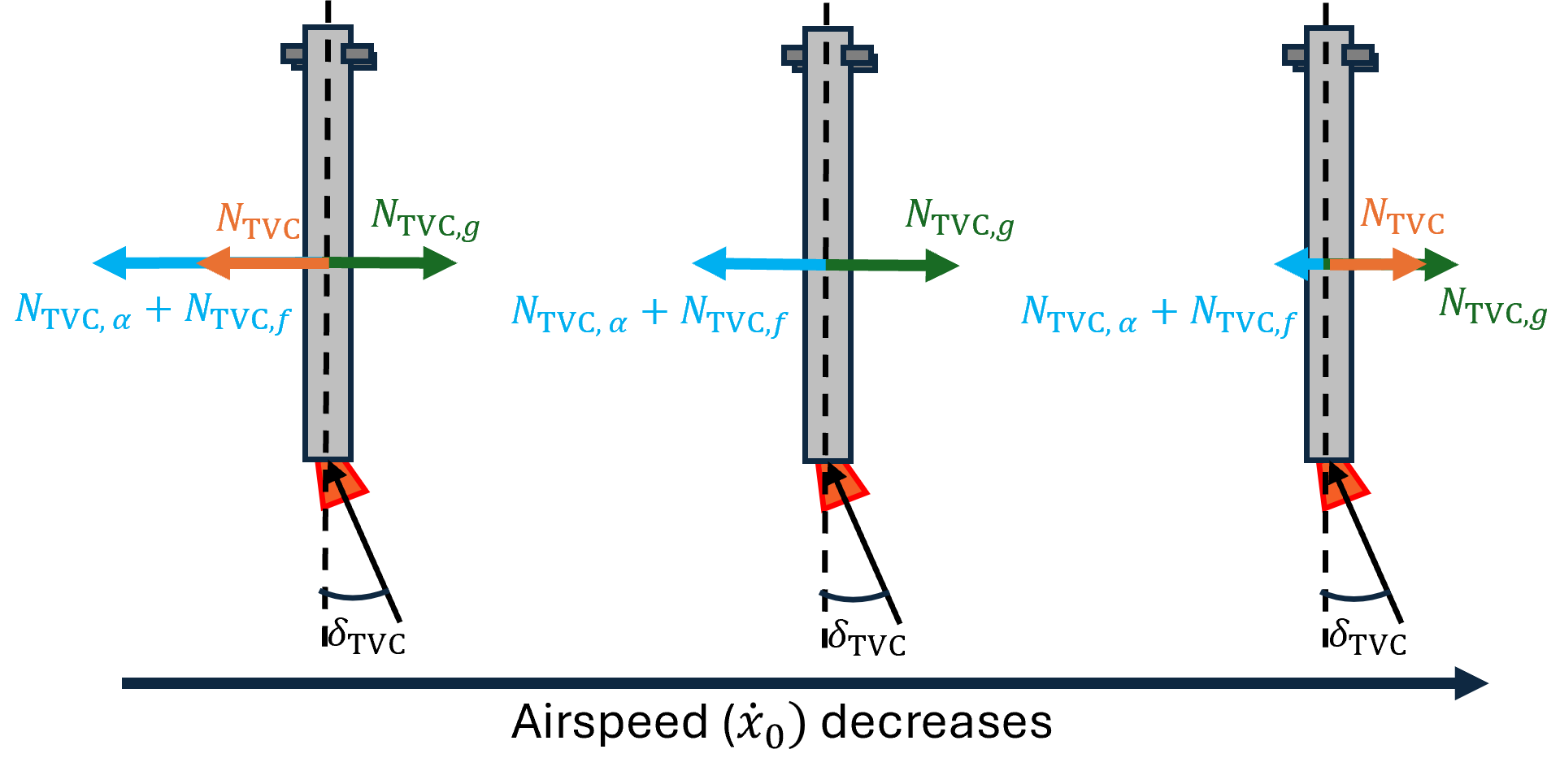}
    \caption{Evolution of the contributions of the $N_{\text{TVC},\, i}$ terms to the low-frequency drift rate dynamics versus airspeed.}
    \label{fig:forcesevolution}
\end{figure}

The transfer function from TVC input to drift rate is analytically derived as:
\begin{equation}
T_{\delta_\text{TVC}\rightarrow \dot{z}} =
\frac{
\dot{x}_0\left(
- Ts^2 + N_{\alpha'}\mu_c + T\mu_{\alpha'} + a_0 m \mu_c
\right)
}{
m \dot{x}_0 s^3 + N_{\alpha'} s^2 - m\mu_{\alpha'} \dot{x}_0 s + a_0 m\mu_{\alpha'}
}
\label{eq:tvc_to_drift_tf}
\end{equation}
The low-frequency behavior is obtained by evaluating the transfer function at \(s=j\omega\) and considering the limit as \(\omega \rightarrow 0\):
\begin{equation}
T_{\delta_\text{TVC}\rightarrow \dot{z}}\Big|_{s = j\omega = 0}=
\frac{
\dot{x}_0\left(
N_{\alpha'}\mu_c + T\mu_{\alpha'} + a_0 m \mu_c
\right)
}{
a_0 m \mu_{\alpha'}
}
\label{eq:dc_gain}
\end{equation}
Since this expression is purely real, the sign of the numerator and of the denominator determines the phase of the system. The denominator maintains a constant sign throughout the flight envelope since, for the entire flight, $m>0$; $a_0=g\sin\theta_0>0$, as the vehicle remains close to vertical ($0<\theta_0<\pi$, Section~\ref{sec:linear_model}); and $\mu_{\alpha'}<0$, since, from Eq.~\eqref{coeffs2}, $\mu_{\alpha'}=(l_\alpha N_\alpha+l_f N_f)/J_N$ with $l_\alpha, l_f>0$ and $N_\alpha, N_f<0$. Hence, the phase behavior depends entirely on the numerator. As \(\dot{x}_0\), \textit{i.e.} the airspeed, does not change sign during flight, the phase sign change depends exclusively on the sign of:
\begin{equation}
N_{\alpha'}\mu_c + T\mu_{\alpha'} + a_0 m \mu_c
\label{eq:sign_condition}
\end{equation}
This expression can be further developed by making use of the relationships given in Eqs.~\eqref{coeffs2} and \eqref{ShortHand}, leading to:
\begin{equation}
\label{condition}
    N_{\alpha}(l_c+l_\alpha)+N_f(l_c+l_f)+a_0ml_c \eqqcolon N_{\text{TVC},\,\alpha} + N_{\text{TVC},\, f} + N_{\text{TVC},\,g} \eqqcolon N_{\text{TVC}}
\end{equation}
The term $N_{\text{TVC},\,\alpha}$ is always negative, and its magnitude tends to decrease over the flight envelope. This behavior can be explained by the fact that the term $N_{\alpha} = Q S_{\mathrm{ref}} C_{N_{\alpha}}$ is always negative and decreases in magnitude due primarily to the reduction in dynamic pressure\footnote{Note that the variation of $C_{N_{\alpha}}$ is not monotonic; however, its contribution is significantly smaller than that of the dynamic pressure.}. The moment arms $l_c$ and $l_{\alpha}$ are always positive and exhibit bounded variations during flight: $l_c$ decreases monotonically, while $l_{\alpha}$ is non-monotonic over most of the trajectory, becoming monotonically decreasing in the second half of the flight envelope.

\par A similar analysis can be performed for the term $N_{\mathrm{TVC},\,f}$. This term is also always negative, and its magnitude decreases over the flight envelope, although not monotonically, as it will be shown later. Regarding the term $N_{\mathrm{TVC},\,g}$, the behavior is similar, with one key difference: this term is always positive. Throughout the flight envelope, its magnitude slightly decreases in a monotonic fashion, since both the mass and the moment arm $l_c$ decrease during flight.

\begin{figure}[t!]
\centering
\begin{subfigure}{0.48\textwidth}
    \centering
    \includegraphics[width=\textwidth]{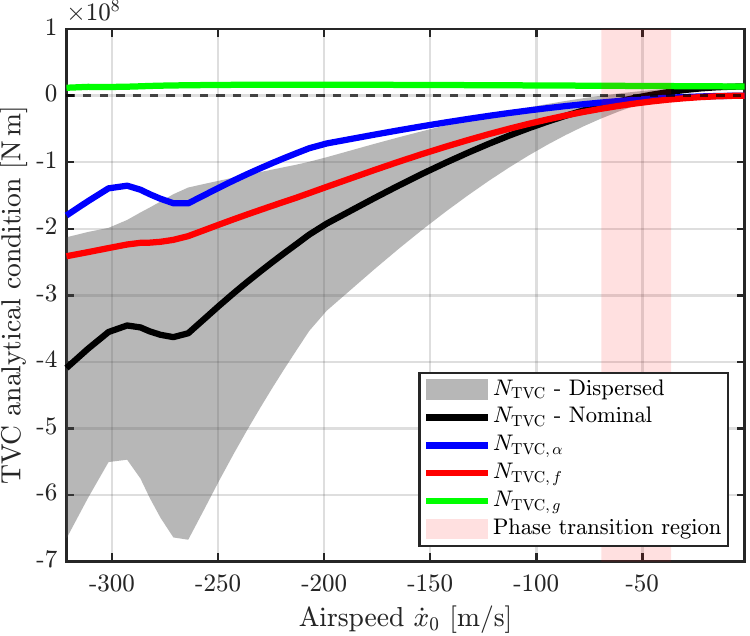}
    \caption{Full envelope.}
    \label{transitionnominal}
\end{subfigure}
\hfill
\begin{subfigure}{0.48\textwidth}
    \centering
    \includegraphics[width=\textwidth]{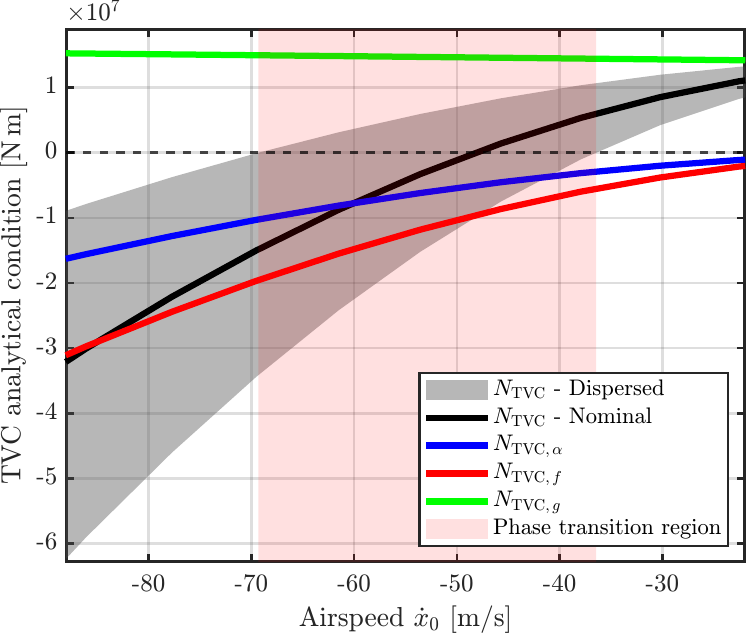}
    \caption{Enlarged view of the transition region.}
    \label{transitionenlarged}
\end{subfigure}
\caption{Analysis of Eq.~\eqref{condition} under nominal and dispersed conditions.}
\label{fig:FPhaseTransitionRegion}
\end{figure}

\par Overall, for the RLV configuration considered in this work, $N_{\mathrm{TVC},\,\alpha}$ and $N_{\mathrm{TVC},\,f}$ are dominant at the beginning of the trajectory and effectively determine the sign of Eq.~\eqref{condition}. However, as velocity decreases, a point is reached where the term $N_{\mathrm{TVC},\,g}$ becomes dominant, leading to a change in the sign of the overall expression; this process is illustrated in Fig.~\ref{fig:forcesevolution}. In addition, Fig.~\ref{fig:FPhaseTransitionRegion} illustrates the evolution of the individual contributions $N_{\mathrm{TVC},\,\alpha}$, $N_{\mathrm{TVC},\,f}$, and $N_{\mathrm{TVC},\,g}$ over the flight envelope under nominal conditions, together with the resulting complete expression ($N_{\mathrm{TVC}}$) evaluated for both nominal and dispersed cases. The dispersed results are obtained via a corner-case analysis, in which all combinations of uncertain parameters at their extreme values are considered in order to characterize the uncertain evolution of Eq.~\eqref{condition}.

\par Figure~\ref{fig:FPhaseTransitionRegion} confirms the previously identified trends in the magnitude and sign of the individual $N_{\mathrm{TVC},i}$ contributions and allows the transition region associated with Eq.~\eqref{condition} to be clearly identified. Under nominal conditions, the transition occurs close to touchdown, at $\dot{x}_0\approx-48.1\,\mathrm{m/s}$, whereas under uncertain conditions the transition airspeed exhibits significant variability. In particular, the sign change is observed to occur over a velocity range of approximately $-69.3\,\mathrm{m/s}$ to $-36.4\,\mathrm{m/s}$ for this specific RLV. Along the nominal trajectory, this interval corresponds approximately to a time window between $31\,\mathrm{s}$ and $35\,\mathrm{s}$, with touchdown occurring at approximately $40\,\mathrm{s}$.

\subsection{Underlying Mechanism - Fins}
\label{Fins_PhaseChange}
\par The transfer function from fin input to drift rate is analytically derived as:
\begin{equation}
T_{\delta_\text{FIN}\rightarrow \dot{z}} = -\frac{\dot{x}_0 \left(N_f s^2 + N_{\alpha'} \mu_{f} - N_{f} \mu_{\alpha'} + a_0 m \mu_{f}\right)}
{m \dot{x}_0 s^3 + N_{\alpha'} s^2 - m \mu_{\alpha'} \dot{x}_0 s + a_0 m \mu_{\alpha'}}
\label{eq:finstodrifttf}
\end{equation}
The low-frequency behavior is obtained by evaluating the transfer function at $s=j\omega$ and considering the limit as $\omega\rightarrow0$:
\begin{equation}
T_{\delta_\text{FIN}\rightarrow \dot{z}}\Big|_{s = j\omega = 0} = -\frac{\dot{x}_0 \left(N_{\alpha'} \mu_{f} - N_{f} \mu_{\alpha'} + a_0 m \mu_{f}\right)}
{a_0 m \mu_{\alpha'}}
\label{eq:dc_gain_fins}
\end{equation}
Similar to the TVC case, this expression is purely real, and for the same considerations as before, the phase shift can be studied by evaluating the sign of the following expression:
\begin{equation}
 N_{\alpha'} \mu_{f} - N_{f} \mu_{\alpha'} + a_0 m \mu_{f}
 \label{tobereferednext}
\end{equation}
Using the relationships of Eqs.~\eqref{coeffs2} and \eqref{ShortHand}, Eq.~\eqref{tobereferednext} can be developed to:
\begin{equation}
    N_\alpha (l_f-l_\alpha) +a_0ml_f \eqqcolon N_{\text{FIN},\, \alpha}+N_{\text{FIN},\, g} \eqqcolon N_{\text{FIN}}
    \label{conditionfins}
\end{equation}
Similarly to Eq.~\eqref{condition}, the aerodynamic term ($N_{\text{FIN},\, \alpha}$) is always negative over the flight envelope, whereas the gravity-related term ($N_{\text{FIN},\,g}$) is always positive. The phase condition is therefore governed by the competition between these two contributions.

\begin{figure}[t!]
\centering
\begin{subfigure}{0.48\textwidth}
    \centering
    \includegraphics[width=\textwidth]{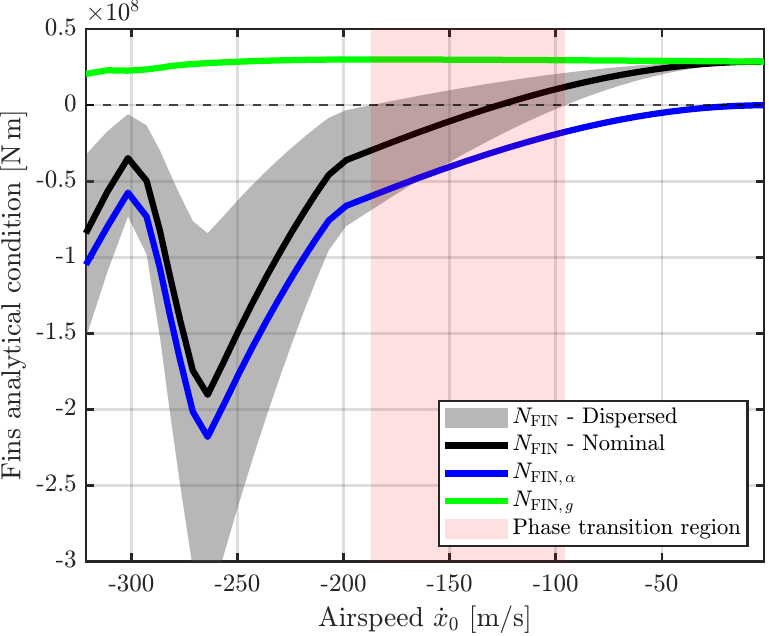}
    \caption{Full envelope.}
    \label{transitionnominalfins}
\end{subfigure}
\hfill
\begin{subfigure}{0.48\textwidth}
    \centering
    \includegraphics[width=\textwidth]{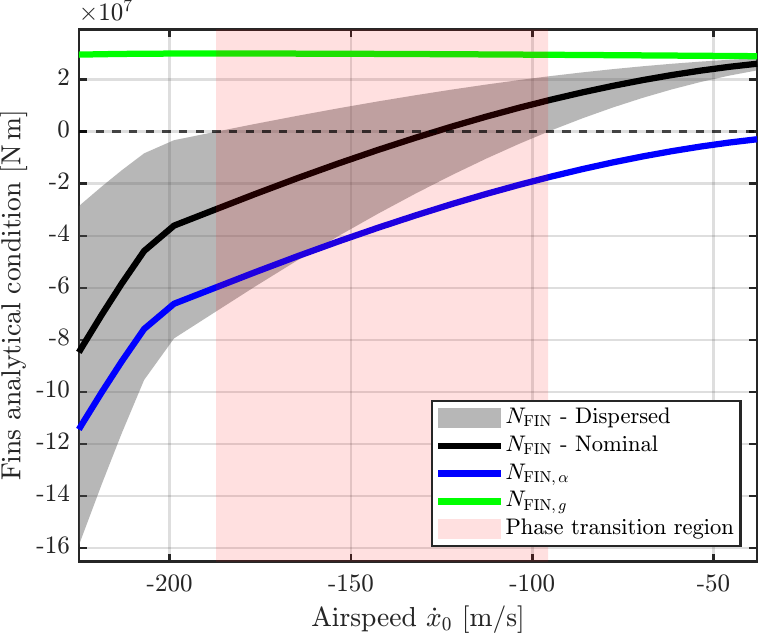}
    \caption{Enlarged view of the transition region.}
    \label{transitionenlargedfins}
\end{subfigure}
\caption{Analysis of Eq.~\eqref{conditionfins} under nominal and dispersed conditions.}
\label{fig:FinsPhaseTransitionRegion}
\end{figure}

Figure~\ref{fig:FinsPhaseTransitionRegion} illustrates the evolution of Eq.~\eqref{conditionfins} over the flight envelope under nominal and dispersed conditions. The overall behavior is consistent with the mechanism observed for the TVC input; however, a key difference is that the transition occurs significantly earlier in the trajectory, at $\dot{x}_0\approx-126.8\,\mathrm{m/s}$ under nominal conditions. Furthermore, the associated transition region exhibits a higher degree of uncertainty, spanning approximately $-187.2\,\mathrm{m/s}$ to $-95.6\,\mathrm{m/s}$, which, along the nominal trajectory, corresponds approximately to a time window between $19\,\mathrm{s}$ and $28\,\mathrm{s}$.

\subsection{Underlying Mechanism - Control Allocation}
\label{Alloc_PhaseChange}
\begin{figure}[t!]
\centering
\begin{subfigure}{0.48\textwidth}
    \centering
    \includegraphics[width=\textwidth]{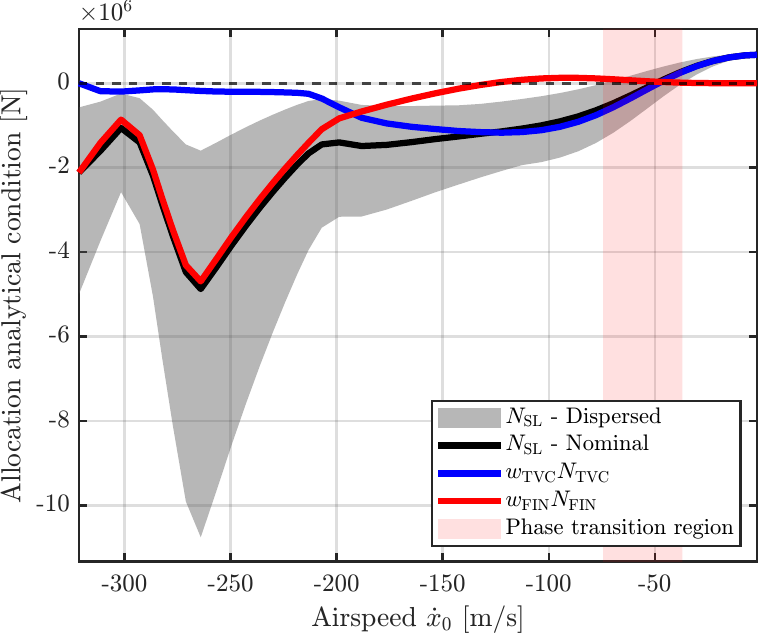}
    \caption{Full envelope.}
\end{subfigure}
\hfill
\begin{subfigure}{0.48\textwidth}
    \centering
    \includegraphics[width=\textwidth]{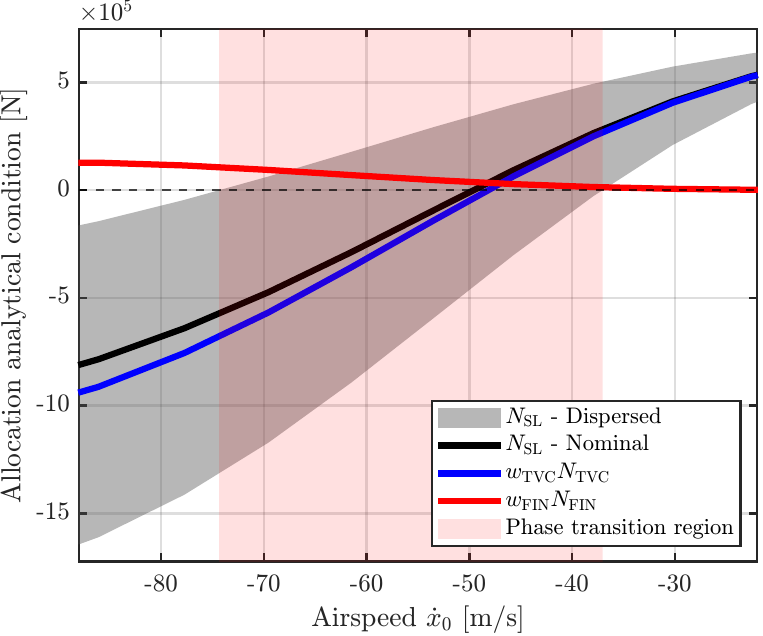}
    \caption{Enlarged view of the transition region.}
\end{subfigure}
\caption{Analysis of Eq.~\eqref{conditionalloc} under nominal and dispersed conditions.}
\label{fig:AllocPhaseTransitionRegion}
\end{figure}

Consider now a single control command $u$ distributed to both actuators through a
control allocation, $[\delta_{\mathrm{TVC}},\ \delta_{\mathrm{FIN}}]^T=[B_1,\ B_2]^T u$,
as in the single-loop INDI configuration of \cite{Diz_2026_esagnc}. By linearity, the
transfer function from $u$ to the drift rate is
$T_{u\rightarrow\dot{z}}=B_1T_{\delta_\text{TVC}\rightarrow\dot{z}}+B_2T_{\delta_\text{FIN}\rightarrow\dot{z}}$,
and, from Eqs.~\eqref{eq:dc_gain} and~\eqref{eq:dc_gain_fins} together with the
simplifications leading to Eqs.~\eqref{condition} and~\eqref{conditionfins}, its
low-frequency gain is
\begin{equation}
    T_{u\rightarrow\dot{z}}\Big|_{s=j\omega=0}=-\frac{\dot{x}_0\,N_{\text{SL}}}{a_0 m\mu_{\alpha'}},
    \label{eq:dc_gain_alloc}
\end{equation}
where $N_{\text{SL}}$ is a weighted combination of the TVC and fin conditions of Eqs.~\eqref{condition} and~\eqref{conditionfins},
\begin{equation}
    N_{\text{SL}} \coloneqq w_{\text{TVC}}N_{\text{TVC}} + w_{\text{FIN}}N_{\text{FIN}},
    \qquad
    w_{\text{TVC}} \coloneqq -\frac{B_1T}{J_N},
    \qquad
    w_{\text{FIN}} \coloneqq \frac{B_2N_f}{J_N}.
    \label{conditionalloc}
\end{equation}
As in the previous cases, the denominator maintains a constant sign, so the phase transition of the allocated input is governed by the sign of $N_{\text{SL}}$. Assuming an allocation such as the pseudo-inverse allocation of \cite{Diz_2026_esagnc}, where $B_1,B_2<0$ (see Eq.~\eqref{eq:B_signs}), both weights are positive, since $T>0$ and $N_f<0$. Consequently, $w_{\text{TVC}}N_{\text{TVC}}$ and $w_{\text{FIN}}N_{\text{FIN}}$ have the same signs as $N_{\text{TVC}}$ and $N_{\text{FIN}}$, respectively. Before the fin transition, $N_{\text{FIN}}$ and $N_{\text{TVC}}$ are both negative, so $N_{\text{SL}}<0$. After the TVC transition, both are positive, so $N_{\text{SL}}>0$. Between the two transitions, $N_{\text{FIN}}>0$ while $N_{\text{TVC}}<0$, and the sign of $N_{\text{SL}}$ depends on which weighted term dominates. The phase transition of the allocated input therefore necessarily occurs between the fin and TVC transitions, at a location set by the allocation weights. For the pseudo-inverse allocation of \cite{Diz_2026_esagnc}, Fig.~\ref{fig:AllocPhaseTransitionRegion} shows that this occurs at $\dot{x}_0\approx-49.6\,\mathrm{m/s}$ along the nominal trajectory, close to the TVC transition. This is the transition that required the gain sign changes of the single-loop controller in~\cite{Diz_2026_esagnc}. Under dispersed conditions, the transition region spans approximately $-74.4\,\mathrm{m/s}$ to $-37.0\,\mathrm{m/s}$, which, along the nominal trajectory, corresponds approximately to a time window between $30\,\mathrm{s}$ and $35\,\mathrm{s}$.

\subsection{Discussion}
\label{sec:discussion} 
Physically, the observed phase transition can be interpreted as the result of a change in the dominant physical mechanism governing lateral drift dynamics along the flight envelope. In the high dynamic pressure regime, the system response is primarily driven by aerodynamic contributions, which therefore define the input–output relationship. As airspeed decreases, aerodynamic authority progressively weakens, leading to a reduction in the relative influence of these terms. This results in a transition toward a regime in which the term associated with $a_0 = g\sin\theta_0$ becomes dominant. This term represents the projection of gravity along the vehicle’s attitude and introduces a kinematic coupling between attitude and drift (see Eq.~\eqref{linEOM}). In this regime, the system behavior resembles that of an inverted pendulum, where the motion is governed less by aerodynamic forces and more by the coupling between attitude and gravity-induced acceleration. The phase inversion observed in both channels, and consequently in the allocated input, occurs at the boundary between these two regimes, where aerodynamic and gravity-driven contributions are of comparable magnitude and the net effective gain changes sign.
\par This transition can also be interpreted through the zeros of the input-to-drift-rate transfer functions. Computing the zeros of Eqs.~\eqref{eq:tvc_to_drift_tf} and~\eqref{eq:finstodrifttf}, using the relationships of Eq.~\eqref{coeffs2}, yields $s=\pm\sqrt{N_{\text{TVC}}/J_N}$ for the TVC input and $s=\pm\sqrt{-N_{\text{FIN}}/J_N}$ for the fin input. The transition corresponds to both zeros reaching the origin, where the low-frequency gain of Eqs.~\eqref{eq:dc_gain} and~\eqref{eq:dc_gain_fins} vanishes. For the TVC input, an imaginary pair becomes real, so that the drift channel becomes non-minimum phase (NMP) after the transition, while the fin input exhibits the reverse behavior. Such zeros are well-known control limitations~\cite{skogestad2005,seron1997fundamental}: a right-half-plane zero at $s=z$ limits the achievable bandwidth to approximately $z/2$. Since in both cases the NMP zeros approach the origin near the transition, the attainable drift and drift rate control bandwidth through either actuator alone tends to zero in its vicinity.
\par Most importantly, the previous analysis shows that the observed transition is fundamentally \textit{a property of the vehicle dynamics rather than of the adopted control architecture.} Moreover, the transition is highly sensitive to modeling uncertainties, as demonstrated previously, which can significantly affect the input-output characteristics near the phase inversion and consequently pose a challenge when lateral position or drift is actively regulated. In Section~\ref{sec:indi}, an INDI-based control architecture capable of robustly handling this transition by using both actuators jointly, while still regulating drift and drift rate errors to zero, will be introduced.
\par Finally, from a design perspective, these results are particularly relevant, as the derived expressions (Eqs.~\eqref{condition}, \eqref{conditionfins}, and~\eqref{conditionalloc}) provide direct insight into the conditions governing the transition. In particular, they make explicit the dependence on aerodynamic, geometric, and mass-related parameters. \textit{These equations can be exploited at the vehicle and mission design level to influence the location of the transition point}, for instance by imposing constraints on mass and geometry during early design stages. Such adjustments can be used to shift the sign inversion away from critical flight phases.

\section{Reconfigurable Incremental Nonlinear Dynamic Inversion}
\label{sec:indi}
This section is divided into five subsections. Section~\ref{sec:indi_fundamentals}
provides a concise theoretical introduction to INDI, largely reproduced from
\cite{JGCD_Diz}, a previous work by the same authors, to ensure that the paper
remains self-contained. Section~\ref{Application} presents the application of
the INDI control law to lateral position and attitude control for the RLV powered landing problem, in both a double-loop and a reconfigured single-loop configuration, while the tuning of the control law using
$\mathcal{H}_\infty$ Closed Loop Shaping is discussed in
Section~\ref{indicontrollertuning}. Section~\ref{INDIPhaseMechanism} provides
a mathematical analysis of how the proposed INDI formulation inherently
handles the phase-transition mechanism described in the previous section.
Finally, Section~\ref{InternalDynamics} derives and analyzes the internal
dynamics introduced by the feedback-linearization process, including their
stability properties.
\subsection{Fundamentals}
\label{sec:indi_fundamentals}
For a particular class of nonlinear systems, the underlying idea behind Incremental Nonlinear Dynamic Inversion (INDI) is to `\textit{cancel out}' the system's nonlinearities through feedback, yielding a closed loop system with linear input-output behavior under the assumption that the inversion is exact. In the context of flight control, for the so-called feedback-linearized system a single outer-loop linear controller can be used to specify the desired performance across the entire flight envelope, thereby eliminating, or at least significantly reducing, the need for gain-scheduling~\cite{pollack,milz2026dynamic}. To illustrate this, consider a multi-input-multi-output (MIMO) nonlinear system, affine in its $m$ inputs, with $n$ states, and the same number of outputs as inputs. The latter assumption is made, without loss of generality~\cite{acquatella2020guidance}, to avoid control-allocation complexities. Such a system can be described by:
\begin{equation}
    \begin{aligned}
    \dot{x} &= f(x) + g(x)u \\
    y &= h(x)
    \end{aligned}
\end{equation}
where $u\in \mathbb{R}^m$ represents the control input vector, $x\in \mathbb{R}^n$ represents the state vector, and $y \in \mathbb{R}^m$ is the output vector. The function $g(x) \in \mathbb{R}^{n\times m}$ is a matrix whose columns are assumed to be smooth vector fields, while $f(x) \in \mathbb{R}^n$ and $h(x) \in \mathbb{R}^m$ are also assumed to be smooth vector fields. Denoting the system's relative degree vector as $\rho = [\rho_1, \dots,\rho_m]^T$\footnote{{The relative degree of an output is the number of times it must be differentiated before the control input appears explicitly.}}, and collecting all differentiated outputs $y_i$ results in $m$ equations in the form of \cite{khalil2002nonlinear,slotine1991applied}:
\begin{equation}
\label{transf}
y^{(\rho)} \triangleq a(x) + \mathcal{B}(x)u, \quad \text{where} \quad
a(x) =
\begin{bmatrix}
\mathcal{L}_f^{\rho_1} h_1(x) \\
\vdots \\
\mathcal{L}_f^{\rho_m} h_m(x)
\end{bmatrix},
\quad
\mathcal{B}(x) =
\begin{bmatrix}
\mathcal{L}_{g_1} \mathcal{L}_f^{\rho_1 - 1} h_1(x) & \cdots & \mathcal{L}_{g_m} \mathcal{L}_f^{\rho_1 - 1} h_1(x) \\
\vdots & \ddots & \vdots \\
\mathcal{L}_{g_1} \mathcal{L}_f^{\rho_m - 1} h_m(x) & \cdots & \mathcal{L}_{g_m} \mathcal{L}_f^{\rho_m - 1} h_m(x)
\end{bmatrix}
\end{equation}
where $\mathcal{L}_f^kh_i(x)$ and $\mathcal{L}_{g_i}\mathcal{L}_f^kh_i(x)$ are recursive Lie derivatives of the function $h_i(x)$ along the vector fields $f(x)$ and $g(x)$ \cite{khalil2002nonlinear}. 
\par To derive the INDI law, consider the following incremental variables given by $\Delta{y}^{(\rho)} = {y}^{(\rho)} - {y}^{(\rho)}_0$, $\Delta x = x - x_0$, and $\Delta u = u - u_0$, which represent the incremental output derivative, incremental state, and incremental control input, respectively. A first-order Taylor series expansion of ${y}^{(\rho)}$ with respect to the delay $\lambda$ is then performed to express the system dynamics in incremental form \cite{bacon2000nonlinear,bacon2001reconfigurable,sieberling2010robust,INDI_Simplicio2025,AcquatellaPhD, xuerei}:
\begin{equation}
\begin{aligned}
       y^{(\rho)} &= y^{(\rho)}_0 +\mathcal{B}(x_0)\Delta u + N_1(x,\lambda) \quad\text{where } 
       N_1(x,\lambda) = \frac{\partial}{\partial x} \left[f(x)+g(x)u \right]\bigg|_{\substack{x=x_0 \\ u = u_0}} \Delta x + \text{H.O.T.}
\end{aligned}
    \label{IncrementalSystem1}
\end{equation}
Assuming that the time delay $\lambda$ is sufficiently small, it can be assumed as well that $\Delta x= x-x_0 \approx0$ \cite{INDI_Simplicio2025} 
and, by neglecting the higher-order terms (H.O.T.) in Eq.~\eqref{IncrementalSystem1} (\textit{i.e.}, $N_1(x,\lambda)\approx 0$), the incremental system reduces to \cite{bacon2000nonlinear,bacon2001reconfigurable,sieberling2010robust,INDI_Simplicio2025,AcquatellaPhD}:
\begin{equation}
    y^{(\rho)} \approx y^{(\rho)}_0+\mathcal{B}(x_0)\Delta u \label{IncrementalSystem}
\end{equation}
Finally, assuming that $\mathcal{B}(x_0)$ is invertible in the relevant domain, NDI can be applied to the incremental system. That is, the incremental control input $\Delta u$ is computed to cancel the effects of ${y}^{(\rho)}_0$ and $\mathcal{B}(x_0)$ in Eq.~\eqref{IncrementalSystem} and
the full INDI control input is then obtained by:
\begin{equation}
    u_\text{INDI} = u_0 + \hat{\mathcal{B}}^{-1}(x_0)[\nu-y^{(\rho)}_0]
    \label{indilaw}
\end{equation}
The INDI formulation does not require knowledge of the full system dynamics $a(x)$ but only of the control-related terms. This significantly reduces sensitivity to modeling errors and facilitates implementation. The resulting control input instead relies on the previous actuation command $u_0$, the measured or estimated output derivative $y^{(\rho)}_0$, and  $\hat{\mathcal{B}}^{-1}(x_0)$. Assuming perfect measurement or estimation of these quantities, the application of the INDI control law to the system in Eq.~\eqref{IncrementalSystem} yields ${y}^{(\rho)} = \nu$ where $\nu$ is the virtual control input; meaning that the implemented control law cancels the nonlinearities and decouples the input–output channels. As a result, each virtual input $\nu_i$ affects only the corresponding output $y_i$, enabling independent controller design for each channel.

\subsection{Application to the Reusable Launch Vehicle }
\label{Application}

\begin{figure}[t!]
    \centering
    \includegraphics[
        width=\linewidth,
        trim={0cm 0cm 0cm 15cm},
        clip
    ]{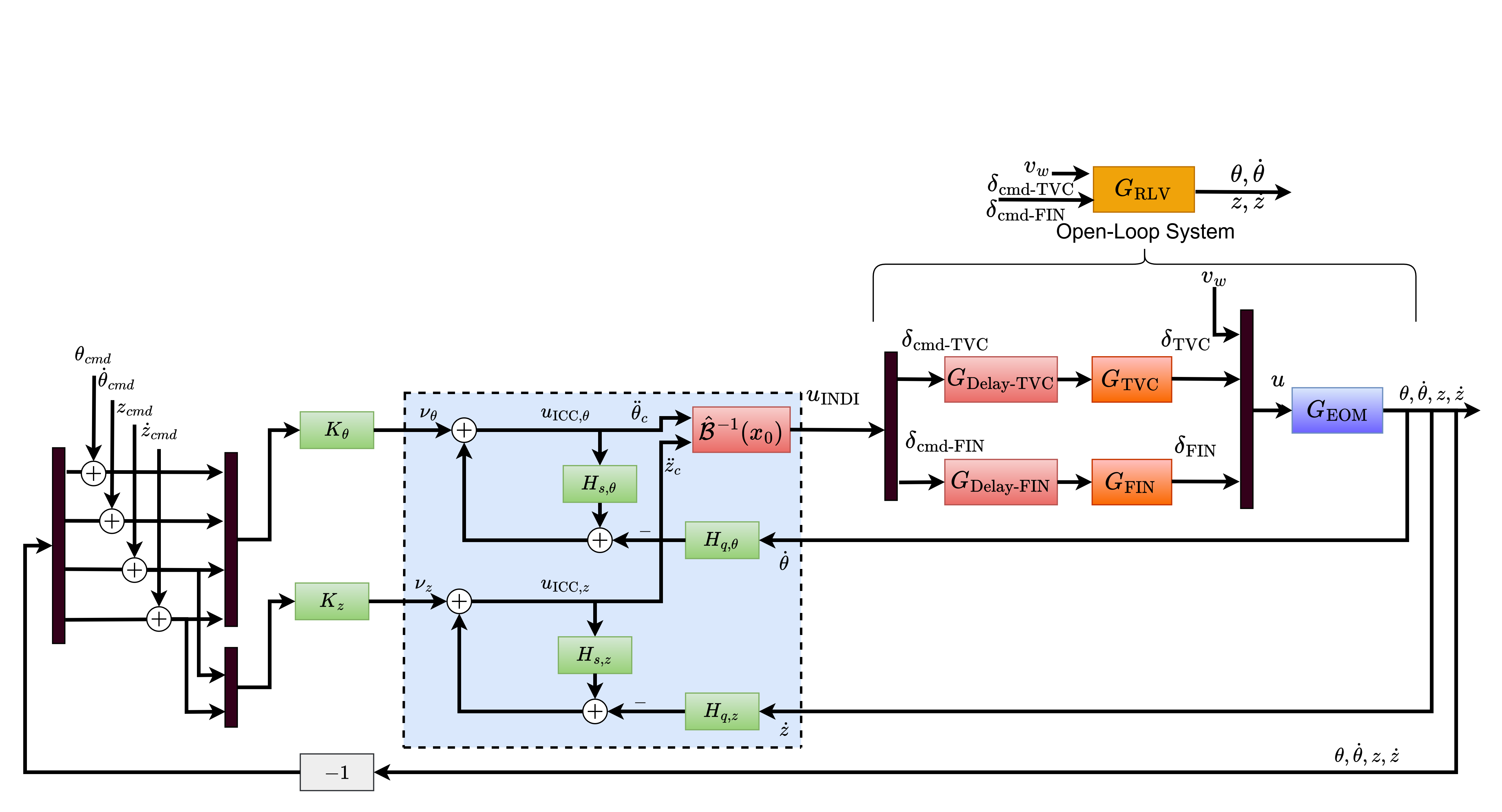}
    \caption{Block-diagram of the double-loop INDI-based control law.}
    \label{fig:InversionScheme}
\end{figure}

The state vector is defined as:
\begin{equation}
x \coloneqq
\begin{bmatrix}
    \theta & q & z & \dot{z}
\end{bmatrix}^{T}
=
\begin{bmatrix}
    x_1 & x_2 & x_3 & x_4
\end{bmatrix}^{T}.
\label{statedef}
\end{equation}
Selecting pitch angle and drift as the controlled outputs,
\begin{equation}
    y = h(x) =
    \begin{bmatrix}
        \theta & z
    \end{bmatrix}^{T},
\end{equation}
the INDI control law can be obtained by substituting the corresponding
launch-vehicle quantities into Eq.~\eqref{indilaw}, yielding
\begin{equation}
\begin{aligned}
u_\text{INDI}
&=
\begin{bmatrix}
    u_{\text{INDI,TVC}} \\
    u_{\text{INDI,FIN}}
\end{bmatrix} =
\begin{bmatrix}
    u_{\text{TVC},0} \\
    u_{\text{FIN},0}
\end{bmatrix}
+
\hat{\mathcal{B}}^{-1}(x_0)
\left(
\begin{bmatrix}
    \nu_\theta \\
    \nu_z
\end{bmatrix}
-
\begin{bmatrix}
    \dot{q}_0 \\
    \ddot{z}_0
\end{bmatrix}
\right) =
\hat{\mathcal{B}}^{-1}(x_0)
\left(
\begin{bmatrix}
    \nu_\theta \\
    \nu_z
\end{bmatrix}
-
\begin{bmatrix}
    \dot{q}_0 \\
    \ddot{z}_0
\end{bmatrix}
+
\hat{\mathcal{B}}(x_0)
\begin{bmatrix}
    u_{\text{TVC},0} \\
    u_{\text{FIN},0}
\end{bmatrix}
\right)
\end{aligned}
\end{equation}
where quantities with the subscript $0$ denote $\lambda$-time-delayed
variables. The inversion loop effectively decouples the attitude and drift
dynamics, allowing the outer-loops to regulate them independently. Two
parallel outer-loop control laws generate the virtual commands
$\nu_\theta$ and $\nu_z$. These controllers, denoted by $K_\theta$ and
$K_z$, respectively, are a double proportional-derivative (PD) controller acting
on pitch angle and drift, and a PD controller acting on drift, as illustrated in Fig.~\ref{fig:InversionScheme}. This structure follows common practice in launch vehicle control, where PD-type controllers are predominantly employed: during ascent, typically one PD acting on drift and one on pitch angle~\cite{Navarro2019,simplicio2019}, and during powered descent, four PDs, two for the TVC and two for the fin commands~\cite{guadagnini2025end}. Here, one PD was initially assigned to each virtual input; an additional PD on drift was then included in the attitude loop to generate angular acceleration commands more compatible with those of the position loop, thereby preventing actuator saturation.

\par In this implementation, the term
$\hat{\mathcal{B}}(x_0)
\begin{bmatrix}
u_{\text{TVC},0}\\
u_{\text{FIN},0}
\end{bmatrix}$
is fed back through the Internal Control Command (ICC) signals
$\begin{bmatrix}
u_{\text{ICC},\theta}\\
u_{\text{ICC},z}
\end{bmatrix}$, following \cite{pollack,Diz_2026_esagnc}.
The relation between the control input and the ICC signals is given by:
\begin{equation}
\begin{bmatrix}
u_{{\text{TVC},0}}\\
u_{{\text{FIN},0}}
\end{bmatrix}
=
G_\text{act}G_\text{delay}
\begin{bmatrix}
u_{\text{cmd,TVC},0}\\
u_{\text{cmd,FIN},0}
\end{bmatrix}
=
G_\text{act}G_\text{delay}
\hat{\mathcal{B}}^{-1}(x_0)
\begin{bmatrix}
u_{\text{ICC},\theta,0}\\
u_{\text{ICC},z,0}
\end{bmatrix}
\end{equation}
where $G_\text{act}$ and $G_\text{delay}$ represent the actuator dynamics
and delays, respectively. Assuming ideal actuator dynamics and negligible
delays,
\begin{equation}
\begin{bmatrix}
u_{\text{TVC},0}\\
u_{\text{FIN},0}
\end{bmatrix}
\approx
\begin{bmatrix}
u_{\text{cmd,TVC},0}\\
u_{\text{cmd,FIN},0}
\end{bmatrix}
=
\hat{\mathcal{B}}^{-1}(x_0)
\begin{bmatrix}
u_{\text{ICC},\theta,0}\\
u_{\text{ICC},z,0}
\end{bmatrix}
\quad\Longrightarrow\quad
\begin{bmatrix}
u_{\text{ICC},\theta,0}\\
u_{\text{ICC},z,0}
\end{bmatrix}
\approx
\hat{\mathcal{B}}(x_0)
\begin{bmatrix}
u_{\text{TVC},0}\\
u_{\text{FIN},0}
\end{bmatrix}.
\end{equation}
Consequently, the ICC provides a direct estimate of the feedback quantity
$\hat{\mathcal{B}}(x_0)\begin{bmatrix}
u_{\text{TVC},0}\\
u_{\text{FIN},0}
\end{bmatrix}$ without requiring an explicit model of the
actuator dynamics. Previous studies \cite{pollack,encarnaccao2025tuning} have
shown that this approximation has a negligible impact on the robustness of
the INDI controller. 
\par In the actual implementation, the available ICC signals correspond to
the current sampling instant. They are therefore passed through the
first-order low-pass filters $H_{s,\theta}$ and $H_{s,z}$, which introduce
the necessary delay to synchronize the ICC signals with the estimated accelerations.\footnote{Further details on the synchronization issue
are provided in Chapter~4 of \cite{pollack}.} The angular and lateral accelerations required by the INDI formulation are estimated using first-order derivative filters, denoted by $H_{q,\theta}$ and $H_{q,z}$, respectively\footnote{Although launch vehicles typically carry linear accelerometers, these measurements were intentionally not used in order to ensure a fair comparison with the baseline controller that will be introduced in Section~\ref{sec:pd_controller}.}.
\smallskip\par
The matrix $\hat{\mathcal{B}}^{-1}(x_0)$ maps the commanded angular and lateral accelerations, $(\ddot{\theta}_c,\ddot{z}_c)$, into TVC and fin deflections. At each operating point, it is computed as the inverse of the local nominal control effectiveness matrix. As the vehicle approaches touchdown, however, the aerodynamic effectiveness of the fins decreases with decreasing airspeed, causing the control effectiveness matrix to become increasingly ill-conditioned. Consequently, its inversion produces excessively large actuator commands.

\smallskip\par
\begin{figure}[t!]
\centering
\includegraphics[width=\linewidth, trim = 0cm 0cm 0cm 20cm]{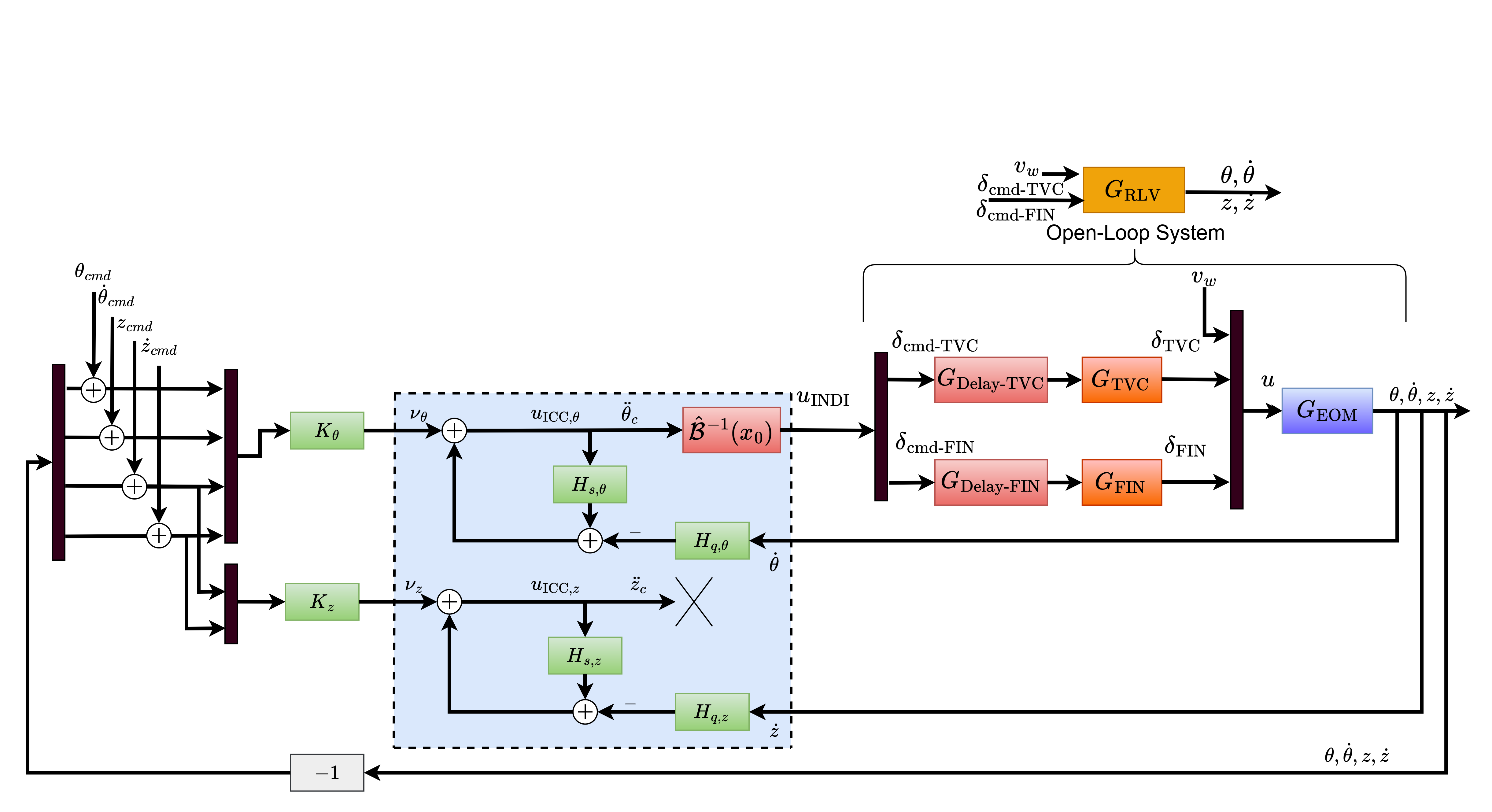}
\caption{Block-diagram of the reconfigured single-loop INDI-based control law.}
\label{fig:InversionSchemeReconf}
\end{figure}

To overcome this limitation, the controller is reconfigured during the final
phase of the landing, as illustrated in Fig.~\ref{fig:InversionSchemeReconf}.
In this configuration, the drift inversion loop is removed, and the explicit
decoupling of attitude and drift is no longer performed\footnote{Avoiding the reconfiguration by retuning the double-loop controller over the final phase of the landing was also attempted. However, the inversion of the increasingly ill-conditioned control effectiveness matrix prevented satisfactory performance and robustness from being achieved.}. Instead, both
attitude and drift are regulated through the attitude inversion loop,
following the approach proposed in \cite{Diz_2026_esagnc}. In the reconfigured controller, $\hat{\mathcal{B}}^{-1}(x_0)$ is replaced by
the Moore-Penrose pseudo-inverse of the row of the control effectiveness
matrix associated with angular acceleration, $[-\mu_c,\ \mu_f]$, which defines the
allocation coefficients $B_1$ and $B_2$ of Section~\ref{Alloc_PhaseChange}. This avoids the ill-conditioning
associated with inversion of the full $\mathcal{B}(x_0)$ matrix while
retaining the contribution of the fins to the control allocation.

\smallskip\par
The proposed reconfiguration is motivated directly by the underlying vehicle
dynamics. As the aerodynamic control effectiveness degrades during the final
phase of landing, the objective of maintaining explicit attitude-drift decoupling becomes increasingly incompatible with the available control authority, since, as $\mu_f$ and $n_f$ tend to zero, the control effectiveness matrix approaches rank one, so that only one output can be inverted independently. The proposed approach therefore relinquishes the decoupling objective and accepts the inherent coupling between attitude and drift, while continuing to regulate both through the attitude inversion loop.
\smallskip 
\par Finally, three remarks are worth noting. First, an alternative to the ICC
approach is to use nominal actuator models to estimate the previous control
signal. Although conceptually simpler, and effective with the double-loop INDI, testing showed that this approach is not compatible with the tuning objectives of the single-loop configuration: when tuned with structured $\mathcal{H}_\infty$ synthesis, no feasible solution satisfying the requirements could be found, presumably because the fixed nominal actuator models restrict the degrees of freedom available to the tuner.
\smallskip

\par Second, the reconfiguration point was selected based on two competing
considerations. On one hand, reconfiguration should be delayed as much as possible,
since the double-loop INDI architecture can accommodate the phase
transition without requiring changes to the outer-loop controller gains,
unlike the single-loop INDI architecture. As discussed in
\cite{Diz_2026_esagnc}, the latter requires changes in the sign of certain
outer-loop gains to account for the phase transition.\footnote{Because the
phase-transition point is uncertain, the required gain sign changes may occur
before or after the actual transition, potentially leading to instability during the transition.} Consequently, delaying the reconfiguration allows the double-loop
architecture to handle the phase transition while avoiding this sensitivity
of the single-loop configuration. On the other hand, the reconfiguration cannot be delayed indefinitely, as the degradation of aerodynamic control effectiveness at low airspeeds causes the required actuator commands to increase significantly. The reconfiguration point was therefore selected heuristically to balance these two considerations, and was chosen to occur gradually between $-37.9$ and $-22.6$~m/s, corresponding to approximately $35$ and $37$~s along the nominal $40$~s landing trajectory. This choice is directly supported by the analysis of Section~\ref{Alloc_PhaseChange}: the reconfiguration starts at approximately the latest point of the phase-transition region of the single-loop configuration ($-37.0\,\mathrm{m/s}$, Fig.~\ref{fig:AllocPhaseTransitionRegion})\footnote{The phase-transition regions are obtained from the input-to-drift-rate transfer functions. For the input-to-drift transfer functions, the result is identical, since the latter are obtained by integrating the drift rate response and therefore differ only by an additional $1/s$ factor, which introduces a constant $-90^\circ$ phase shift without affecting the $180^\circ$ phase inversion.}, and, since it is performed gradually, the single-loop becomes fully active only after the transition has occurred for all considered dispersions.

\par Third, the proposed reconfiguration could potentially be avoided by
employing more advanced control-allocation strategies within the double-loop
INDI. One possible approach is to solve an optimization-based control allocation problem on board~\cite{johansen2013control,bodson2002evaluation}, rather than directly inverting the control effectiveness matrix, thereby preventing the excessively large actuator commands that result from inverting a near-singular matrix. This is left as a potential direction for future work.

\smallskip

\smallskip

\subsection{Controller Tuning}
\label{indicontrollertuning}

    The tuning of the control system aims at optimizing the components represented by the green blocks in Figs.~\ref{fig:InversionScheme} and ~\ref{fig:InversionSchemeReconf}. In this work, $\mathcal{H}_{\infty}$ Closed Loop Shaping is used for controller synthesis following the procedure described next. First, all tunable components are synthesized simultaneously at a flight condition corresponding to an early stage of the flight (in this work, $\dot{x}_0 = -249.6\,\mathrm{m/s}$). Subsequently, the first-order filters ($H_{q,\theta}$, $H_{s,\theta}$, $H_{q,z}$, $H_{s,z}$) are fixed, and $K_{\text{Out}}\coloneqq \{K_\theta,K_z\}$ is recomputed for the remaining flight points\footnote{Note that, even with the INDI-based architecture, more than a single flight point is required to satisfy the stability requirements. The flight points were selected heuristically to ensure that the input disk-based \textit{S-T} loop-at-a-time stability margins remain above $6\,\mathrm{dB}$ and $30^\circ$. These margins were selected because they provide a symmetric measure of robustness to gain and phase variations and account for simultaneous gain and phase perturbations. For a detailed discussion of disk margins and their advantages over classical gain and phase margins, the reader is referred to \cite{Seiler2020DiskMargins}.}. Two operating points are required to satisfy the stability requirements over the nominal flight envelope, while a third one is introduced to synthesize the reconfigured controller used near touchdown. The resulting controllers are gain-scheduled as a function of airspeed, consistent with the scheduling adopted for $\hat{\mathcal{B}}^{-1}(x_0)$. It should be noted that the controller gains and control allocation are held constant outside the scheduling range, using the values from the nearest design point. This avoids extrapolating the gain scheduling beyond the synthesized flight points, where the gain trends are unknown, at the cost of some performance degradation.

\begin{table}[t!]
\centering
\caption{$\mathcal{H}_\infty$ Closed Loop Shaping constraints used for controller design.}
\begin{tabular}{l|ccc|}
\cline{2-4}
\multicolumn{1}{c|}{} &
\multicolumn{3}{c|}{Flight Point (Airspeed in $\mathrm{m/s}$)} \\ \hline

\multicolumn{1}{|c|}{Constraints} &
\multicolumn{1}{c|}{$\dot{x}_0 =  -249.6$} &
\multicolumn{1}{c|}{$\dot{x}_0 =  -37.9$} &
\multicolumn{1}{c|}{$\dot{x}_0 =  -22.6$}\\ \hline

\multicolumn{1}{|l|}{LAT DB Margins} &
\multicolumn{1}{c|}{$2.5\,\mathrm{dB},\,12.5^\circ$} &
\multicolumn{1}{c|}{$6\,\mathrm{dB},\,30^\circ$} &
$6\,\mathrm{dB},\,30^\circ$ \\ \hline

\multicolumn{1}{|l|}{$\|T_{\theta_\mathrm{cmd} \rightarrow \delta_\text{TVC}}\|<W_1$} &
\multicolumn{1}{c|}{$0.17$} &
\multicolumn{1}{c|}{$2$} &
\multicolumn{1}{c|}{$2$} 
 \\ \hline

\multicolumn{1}{|l|}{$\|T_{\theta_\mathrm{cmd} \rightarrow \delta_\text{FIN}}\|<W_2$} &
\multicolumn{1}{c|}{$4$} &
\multicolumn{1}{c|}{$4$} &
\multicolumn{1}{c|}{$4$} 
 \\ \hline

 \multicolumn{1}{|l|}{$\|T_{z_\mathrm{cmd} \rightarrow \delta_\text{TVC}}\|<W_3$} &
\multicolumn{1}{c|}{$0.17$} &
\multicolumn{1}{c|}{$2$} &
\multicolumn{1}{c|}{$2$} 
 \\ \hline

\multicolumn{1}{|l|}{$\|T_{z_\mathrm{cmd} \rightarrow \delta_\text{FIN}}\|<W_4$} &
\multicolumn{1}{c|}{$4$} &
\multicolumn{1}{c|}{$4$} &
\multicolumn{1}{c|}{$4$} 
 \\ \hline

\multicolumn{1}{|l|}{$\|T_{z_{\mathrm{cmd}} \rightarrow z}\|_\infty < W_5$} &
\multicolumn{1}{c|}{$\frac{1.90s}{s + 0.405}$} &
\multicolumn{1}{c|}{$\frac{1.32s}{s + 0.405}$} &
\multicolumn{1}{c|}{$\frac{3.5s}{s + 0.405}$} 
 \\ \hline

\multicolumn{1}{|l|}{$\|T_{\theta_{\mathrm{cmd}} \rightarrow \theta}\|_\infty < W_6$} &
\multicolumn{1}{c|}{$\frac{1.9(2s+0.01)}{s + 0.1}$} &
\multicolumn{1}{c|}{$\frac{0.74(2s+0.01)}{s + 0.1}$} &
\multicolumn{1}{c|}{$\frac{1.75(2s+0.01)}{s + 0.1}$}  \\ \hline

\multicolumn{1}{|l|}{$\|T_{v_w \rightarrow z}\|_\infty < W_7$} &
\multicolumn{1}{c|}{$13.16$} &
\multicolumn{1}{c|}{$25$} &
\multicolumn{1}{c|}{$100$} \\ \hline

\multicolumn{1}{|l|}{$\|T_{v_w \rightarrow \dot{z}}\|_\infty < W_8$} &
\multicolumn{1}{c|}{$5.12$} &
\multicolumn{1}{c|}{$5$} &
\multicolumn{1}{c|}{$100$} \\ \hline

\end{tabular}
\label{HinftuningINDI}
\end{table}

\smallskip
\par The synthesis objectives used for controller design are summarized in Table~\ref{HinftuningINDI}. The entries in the loop-at-a-time (LAT) disk-based (DB) margin row correspond to the minimum required stability margins enforced in the synthesis, while the entries associated with the performance objectives correspond to the weighting functions $W_1$-$W_8$ used at each flight design point. It should be noted that, at the first flight point, the margin requirements are enforced over a set of models generated from the corner cases of all LTI uncertainties  (hence the reduced $2.5\,\mathrm{dB}$, $12.5^\circ$ requirement), as this was found to be necessary to guarantee stability across all parameter combinations.
\smallskip
\par The first four performance objectives, $\|T_{\theta_\mathrm{cmd} \rightarrow \delta_\text{TVC}}\|<W_1$, $\|T_{\theta_\mathrm{cmd} \rightarrow \delta_\text{FIN}}\|<W_2$, $\|T_{z_\mathrm{cmd} \rightarrow \delta_\text{TVC}}\|<W_3$, and $\|T_{z_\mathrm{cmd} \rightarrow \delta_\text{FIN}}\|<W_4$, penalize the commanded TVC and fin deflections, thereby preventing excessive actuator activity. The fifth performance objective, $\|T_{z_{\mathrm{cmd}} \rightarrow z}\|_\infty < W_5$, enforces tracking of lateral position commands, while the sixth, $\|T_{\theta_{\mathrm{cmd}} \rightarrow \theta}\|_\infty < W_6$, enforces pitch angle tracking. They are adjusted to improve tracking while maintaining adequate actuation levels. Finally, the last two objectives, $\|T_{v_w \rightarrow z}\|_\infty < W_7$ and $\|T_{v_w \rightarrow \dot{z}}\|_\infty < W_8$, limit the effect of lateral wind disturbances on drift and drift rate. These weighting functions are relaxed throughout the descent phase (see Table~\ref{HinftuningINDI}) to account for the reduction in control authority at lower airspeeds. It is important to note that the synthesis minimizes the soft goals subject to the hard goals as constraints~\cite{apkarian2014multi}. The objectives required for a successful landing, namely the stability margins, tracking ($W_5$, $W_6$), and disturbance rejection ($W_7$, $W_8$), are enforced as hard constraints, while the actuation objectives ($W_1$-$W_4$) are treated as soft goals, such that the required performance is achieved with minimal actuator usage.

\subsection{Phase Transition Mechanism Effects}
\label{INDIPhaseMechanism}

The phase-transition mechanism identified in Section~\ref{sec:l1} does not impose the same limitations on the proposed double-loop INDI architecture as it does on the single-loop configuration considered in \cite{Diz_2026_esagnc}. In particular, the double-loop INDI remains active through and beyond the phase-transition region without requiring any sign changes in the outer-loop gains, and, as shown by the stability analysis in Section~\ref{Stability Analysis}, remains stable throughout the flight envelope. The reason for this difference lies in how the double-loop architecture separates the attitude and drift dynamics, as discussed next.

\par Recall that, at low frequencies, the sign of the open-loop transfer
functions from the TVC and fin inputs to drift rate is determined by the sign
of:
\begin{equation}
    N_{\alpha}(l_c+l_\alpha)+N_f(l_c+l_f)+a_0ml_c,
    \label{ref1}
\end{equation}
and
\begin{equation}
    N_\alpha (l_f-l_\alpha) +a_0ml_f,
    \label{ref2}
\end{equation}
respectively. In both cases, the expressions are initially negative and
approach zero as the airspeed decreases, eventually becoming positive. The term that enables this sign reversal is the contribution involving $a_0=g\sin(\theta_0)$: if $a_0$ were zero, neither expression would change sign over the considered flight envelope.

\par To understand the origin of this effect, consider the
linearized equation of motion for the lateral acceleration:
\begin{equation}
    \ddot{z} =
    \left(-\frac{N_{\alpha'}}{m}-a_0 \right)\theta
    - \frac{N_{\alpha'}}{m\dot{x}_0}\dot{z}
    -\frac{T}{m}\delta_\text{TVC}
    -\frac{N_f}{m}\delta_\text{FIN}
    + \frac{N_{\alpha'}}{m\dot{x}_0}v_w.
\end{equation}
The $a_0$-dependent term represents a cross-coupling between vehicle attitude
and lateral acceleration; notably, no $a_0$-dependent term appears in the angular acceleration equation. If the channels were fully decoupled, the contribution
responsible for the sign reversal would be canceled and, consequently,
the associated $180^\circ$ phase transition would no longer occur.

\par The proposed double-loop INDI architecture exploits the availability of two inputs to invert the angular and lateral acceleration channels independently. In doing so, the inversion cancels the state-dependent terms of both channels, including the $a_0$-dependent cross-coupling term, resulting, ideally, in the following decoupled input-output relationships:
\begin{equation}
\ddot{\theta} = \nu_\theta,
\qquad
\ddot{z} = \nu_z.
\end{equation}
Consequently, the drift channel seen by the position-control outer-loop is ideally a double integrator, $z/\nu_z = 1/s^2$, whose sign remains fixed throughout the flight envelope. The phase transition therefore does not appear in this channel, and no sign change is required in the outer-loop gains. This holds even though the physical coupling remains present in the vehicle dynamics. Without the inversion, this coupling must instead be accounted for through the controller gains, as in the baseline controller of Section~\ref{sec:pd_controller}.
\par In contrast, in the single-loop configuration of \cite{Diz_2026_esagnc}, attitude and drift are regulated through a single command distributed to both actuators by the control allocation. The attitude-drift coupling remains in the input-output dynamics seen by the outer-loop, whose drift channel undergoes the phase transition governed by $N_{\text{SL}}$ (Eq.~\eqref{conditionalloc}), requiring the corresponding sign changes in the outer-loop controller gains.

\subsection{Internal Dynamics Derivation and Stability Analysis}
\label{InternalDynamics}

When feedback linearization is applied to a system, the resulting dynamics
are divided into two parts: the external dynamics (usually denoted by $\xi$), which are directly influenced by the control inputs, and the internal dynamics (usually denoted by $\eta$), which are uncontrollable through them \cite{JGCD_Diz}. The existence of internal dynamics in a feedback-linearized system is
determined by the relationship between the total relative degree of the
controlled outputs, $\rho_t$, and the system order, $n$. The total relative
degree is defined as $\rho_t=\sum_i \rho_i$, where $\rho_i$ denotes the
relative degree of the $i$th controlled output. When $\rho_t=n$, no internal dynamics are present, whereas for
$\rho_t<n$, the remaining $n-\rho_t$ states constitute the internal dynamics.

For the RLV considered in this work, the system has four states,
$n=4$, given by $(\theta,\dot{\theta},z,\dot{z})$. In the double-loop INDI architecture, both $\theta$ and $z$ are selected as controlled outputs. As can be seen from Eq.~\eqref{CA_final}, both outputs have relative degree two, such that: $\rho_\theta = \rho_z = 2$, and consequently the total relative degree is $\rho_t = \rho_\theta + \rho_z = 4 = n$. Since the total relative degree is equal to the system order, no internal dynamics are present. In contrast, the single-loop INDI architecture considered in this work regulates only the attitude output $\theta$. Its relative degree is two, yielding
\begin{equation}
    \rho_t = \rho_\theta = 2 < n.
\end{equation}
Consequently, the feedback-linearized single-loop INDI system has
$n-\rho_t=2$ internal states. These can be characterized by smooth functions
$\eta_i(x)$ that are invariant along the input vector fields $g_j(x)$, while also satisfying $\eta(0)=0$:
\begin{equation}
    \frac{\partial \eta_i}{\partial x}g_j(x)=0,
    \qquad
    i=1,\ldots,n-\rho_t,\quad j=1,\ldots,m,
    \qquad
    \eta(0)=0.
\label{conditions}
\end{equation}
For the single-loop INDI architecture, the input vector field must account
for the control allocation used to generate the physical TVC and fin commands
from the virtual input. With the allocation coefficients $B_1$ and $B_2$ defined in Section~\ref{Alloc_PhaseChange}, the resulting
effective input vector field is therefore given by:
\begin{equation}
    g_{\mathrm{INDI}}(x)=
    \begin{bmatrix} 0 \\ -\mu_cB_1+\mu_fB_2 \\ 0 \\ -n_cB_1-n_fB_2 \end{bmatrix}.
    \label{gindi}
\end{equation}
For convenience, define the second and fourth components of this vector field as
\begin{equation}
    A \coloneqq -\mu_cB_1+\mu_fB_2,
    \qquad
    C \coloneqq  -n_cB_1-n_fB_2.
    \label{eq:A_definitions}
\end{equation}
Using the general equation defining the internal dynamics (Eq.~\eqref{conditions}), together with the input vector field $g_{\mathrm{INDI}}$ (Eq.~\eqref{gindi}), constraints on the internal states of the single-loop INDI can be obtained:
\begin{equation}
    \eta(0)=0,
    \qquad
    A\frac{\partial\eta_i}{\partial\dot{\theta}}
    +
    C\frac{\partial\eta_i}{\partial\dot{z}}
    =0,
    \qquad
    i=1,\ldots,n-\rho_t.
    \label{eq:internal_dynamics_conditions}
\end{equation}
The choice of $\eta(x)$ is not unique; one possible realization is obtained
by directly selecting as internal states the state variables whose partial
derivatives with respect to the rate states are zero, namely $z$, and by
constructing linear combinations of the remaining state variables that
satisfy the conditions defined above. This yields:
\begin{equation}
    \eta(x) = [\eta_1,\, \eta_2] = [z,\, C\dot{\theta}-A\dot{z}]
\end{equation}
Having defined the internal dynamics, a mapping, $T(x)$, between the original state variables and the external and internal states can now be established:
\begin{equation}
T(x)=
    \left[
    \begin{array}{c}
        \xi(x)\\
        \hline
        \eta(x)
    \end{array}
    \right]
    =
    \left[
    \begin{array}{c}
        \theta\\
        \dot{\theta}\\
        \hline
        z\\
        C\dot{\theta}-A\dot{z}
    \end{array}
    \right]
    =
    \left[
    \begin{array}{c}
        x_1\\
        x_2\\
        \hline
        x_3\\
        Cx_2-Ax_4
    \end{array}
    \right].
    \label{eq:internal_external_state_coordinates}
\end{equation}
As discussed in \cite{JGCD_Diz}, the transformation $T$ is required to be a
diffeomorphism over a neighborhood $\mathcal{D}_0$ of $x$, where
$\mathcal{D}_0 = \{\tilde{x} \in \mathbb{R}^n:\|\tilde{x}-x\|<\epsilon\}$,
because it must be locally invertible, \textit{i.e.},
$T^{-1}(T(x))=x$ for all $x\in\mathcal{D}_0$, and
both $T$ and $T^{-1}$ must be smooth mappings in $\mathbb{R}^n$. These
properties ensure that the nonlinear system remains smooth under the
coordinate transformation. To prove that $T$ is a valid diffeomorphism, observe that:
\begin{equation}
\begin{aligned}
    x &= T^{-1}(T(x)) = \left[x_1,\,x_2,\,x_3,\,x_4\right]^T = \left[\xi_1,\, \xi_2,\,\eta_1,\,\frac{C}{A}\xi_2-\frac{1}{A}\eta_2\right]
\end{aligned}
\end{equation}
which shows that $T$ is locally invertible as long as $A\neq0$. Since $B_1$ and $B_2$ are obtained from the pseudo-inverse of the angular-acceleration row $[-\mu_c,\ \mu_f]$ of the control-effectiveness matrix, $A=-\mu_cB_1+\mu_fB_2=1$ for the nominal model at every operating point. Consequently, $T$ is a valid diffeomorphism throughout the flight envelope, and the internal dynamics can be obtained by differentiating the internal coordinates in Eq.~\eqref{eq:internal_external_state_coordinates}:
\begin{equation}
    \dot{\eta}
    =
    \begin{bmatrix}
        \dot{z}\\
        C\ddot{\theta}-A\ddot{z}
    \end{bmatrix}
    \label{eq:internal_dynamics_derivative}
\end{equation}
Since the internal dynamics will be analyzed locally about the reference trajectory, the linearized dynamics of Eq.~\eqref{linEOM} are used (perturbation notation omitted for readability). Substituting these dynamics into Eq.~\eqref{eq:internal_dynamics_derivative} yields:
\begin{equation}
    \begin{aligned}
        \dot{\eta}_1 &= \dot{z},\\
        \dot{\eta}_2 &=
        C\left(\mu_{\alpha'}\alpha + A\,\ddot{\theta}_c\right)
        -A\left(-n_{\alpha'}\alpha - a_0\theta + C\,\ddot{\theta}_c\right).
    \end{aligned}
    \label{eq:internal_dynamics_substituted}
\end{equation}
Using the definitions of $A$ and $C$ in Eq.~\eqref{eq:A_definitions}, and expressing the resulting dynamics in terms of the external and internal coordinates $\xi$ and $\eta$, yields: 
\begin{equation}
    \begin{aligned}
        \dot{\eta}_1
        &=
        \frac{C}{A}\xi_2-\frac{1}{A}\eta_2,\\
                \dot{\eta}_2
        &=
        C\mu_{\alpha'}\alpha(\xi,\eta)
        +A\left(n_{\alpha'}\alpha(\xi,\eta) + a_0\xi_1\right).
    \end{aligned}
    \label{eq:internal_dynamics_xi_eta}
\end{equation}
Subsequently, the zero dynamics of the system, \textit{i.e.}, the internal-state dynamics
for $\xi=0$, can then be obtained as:
\begin{equation}
    \begin{aligned}
        \dot{\eta}_{0,1}
        &=
        -\frac{1}{A}\eta_{0,2},\\
        \dot{\eta}_{0,2}
        &=
        C\mu_{\alpha'}\alpha(0,\eta_0)
        +An_{\alpha'}\alpha(0,\eta_0).
    \end{aligned}
    \label{eq:zero_dynamics}
\end{equation}
where the subscript $0$ denotes the zero dynamics. Expressing the angle of attack as a function of the internal and external
states and evaluating it for $\xi=0$:
\begin{equation}
    \alpha
    =
    \frac{\dot{z}+\theta\dot{x}_0}{\dot{x}_0}
    =
    \frac{
    \xi_2\frac{C}{A}
    -\eta_2\frac{1}{A}
    +\xi_1\dot{x}_0
    }{\dot{x}_0}
    \xrightarrow{\xi=0}
    -\frac{\eta_{0,2}}{A\dot{x}_0},
    \label{eq:alpha_zero_dynamics}
\end{equation}
Substituting Eq.~\eqref{eq:alpha_zero_dynamics} into
Eq.~\eqref{eq:zero_dynamics} then gives the following representation of the
zero dynamics:
\begin{equation}
    \begin{bmatrix}
        \dot{\eta}_{0,1}\\
        \dot{\eta}_{0,2}
    \end{bmatrix}
    =
    \begin{bmatrix}
        0 & -\dfrac{1}{A}\\[2ex]
        0 & -\dfrac{1}{A\dot{x}_0}
        \left(C\mu_{\alpha'}+An_{\alpha'}\right)
    \end{bmatrix}
    \begin{bmatrix}
        \eta_{0,1}\\
        \eta_{0,2}
    \end{bmatrix}.
    \label{eq:internal_dynamics_matrix}
\end{equation}
The stability of the internal dynamics can be assessed by evaluating the eigenvalues of the system matrix. Since the matrix in Eq.~\eqref{eq:internal_dynamics_matrix} is upper triangular, its eigenvalues are given directly by its diagonal entries:
\begin{equation}
    \lambda_1=0,
    \qquad
    \lambda_2=
    -\frac{1}{A\dot{x}_0}
    \left(C\mu_{\alpha'}+An_{\alpha'}\right).
    \label{eq:internal_dynamics_eigenvalues}
\end{equation}
The zero eigenvalue corresponds to the drift state $\eta_1=z$, which behaves as an integrator and is therefore marginally stable. The stability of the remaining internal dynamics is determined by the sign of $\lambda_2$. Substituting the definitions of $A$ and $C$ from Eq.~\eqref{eq:A_definitions}, together with the definitions of the aerodynamic derivatives, gives
\begin{equation}
\begin{aligned}
    \lambda_2
    &=
    -\frac{1}{A\dot{x}_0 m J_N}
    \left[
    -B_1T\left(N_\alpha(l_c+l_\alpha)+N_f(l_c+l_f)\right)
    +B_2N_\alpha N_f(l_f-l_\alpha)
    \right]\\
    &=
    \frac{B_1T\left(N_\alpha(l_c+l_\alpha)+N_f(l_c+l_f)\right)}
    {A\dot{x}_0mJ_N}
    +
    \frac{N_\alpha N_fB_2(l_\alpha-l_f)}
    {A\dot{x}_0mJ_N}
    \triangleq
    \lambda_{2,\mathrm{TVC}}
    +
    \lambda_{2,\mathrm{FIN}}.
\end{aligned}
\label{eq:lambda_2_decomposition}
\end{equation}
The sign of $\lambda_2$ can be established analytically over the entire flight envelope.
First, since $B_1$ and $B_2$ are obtained from the pseudo-inverse of the
angular-acceleration row $[-\mu_c,\ \mu_f]$, they share the signs of its entries:
\begin{equation}
    B_1=-\frac{\mu_c}{\mu_c^2+\mu_f^2}<0,
    \qquad
    B_2=\frac{\mu_f}{\mu_c^2+\mu_f^2}<0,
    \label{eq:B_signs}
\end{equation}
since $\mu_c=l_cT/J_N>0$ and $\mu_f=l_fN_f/J_N<0$. Second, the term in parentheses in
$\lambda_{2,\mathrm{TVC}}$ is precisely the aerodynamic part of the TVC phase-transition
condition of Eq.~\eqref{condition}, i.e.,
$N_\alpha(l_c+l_\alpha)+N_f(l_c+l_f)=N_{\text{TVC},\alpha}+N_{\text{TVC},f}$,
which was shown in Section~\ref{TVC_PhaseChange} to be negative throughout the flight
envelope. Hence, the numerator of $\lambda_{2,\mathrm{TVC}}$, given by the product of
$B_1<0$, $T>0$, and $N_{\text{TVC},\alpha}+N_{\text{TVC},f}<0$, is positive. Third, in
$\lambda_{2,\mathrm{FIN}}$, both $N_\alpha$ and $N_f$ are negative, so that
$N_\alpha N_f>0$, while $l_\alpha-l_f<0$, since the fins are located well ahead of the center of pressure
($l_f>l_\alpha>0$ throughout the flight envelope).
Together with $B_2<0$, the numerator of $\lambda_{2,\mathrm{FIN}}$ is therefore also
positive. Finally, the common denominator $A\dot{x}_0mJ_N$ is negative, since $A=1$,
$m,J_N>0$, and $\dot{x}_0<0$ during descent. Both $\lambda_{2,\mathrm{TVC}}$ and
$\lambda_{2,\mathrm{FIN}}$ are thus negative and, consequently, \textit{the second internal state is exponentially stable throughout the flight envelope}. Note that this conclusion is not restricted to the pseudo-inverse allocation: it holds for any allocation with $B_1<0$ and $B_2<0$, since $A=-\mu_cB_1+\mu_fB_2>0$ is then also guaranteed. 

\par At low airspeeds, the contribution of the fins becomes negligible ($B_2\approx0$, and hence $A=-\mu_cB_1=1$), and Eq.~\eqref{eq:lambda_2_decomposition} reduces to
\begin{equation}
    \lambda_2
    =
    -\frac{T\left(N_\alpha(l_c+l_\alpha)+N_f(l_c+l_f)\right)}
    {\mu_c\dot{x}_0mJ_N}
    <0,
    \label{eq:lambda_2_low_speed}
\end{equation}
which, for $N_f\rightarrow0$, \textit{i.e.}, removing the residual fin normal force that remains in Eq.~\eqref{eq:lambda_2_low_speed} and thus reducing the vehicle to a TVC-only configuration, coincides with the second internal-dynamics eigenvalue of the rigid-body ascending launch vehicle in \cite{JGCD_Diz}.

\begin{figure}[t!]
    \centering
    \includegraphics[width=0.6\linewidth]{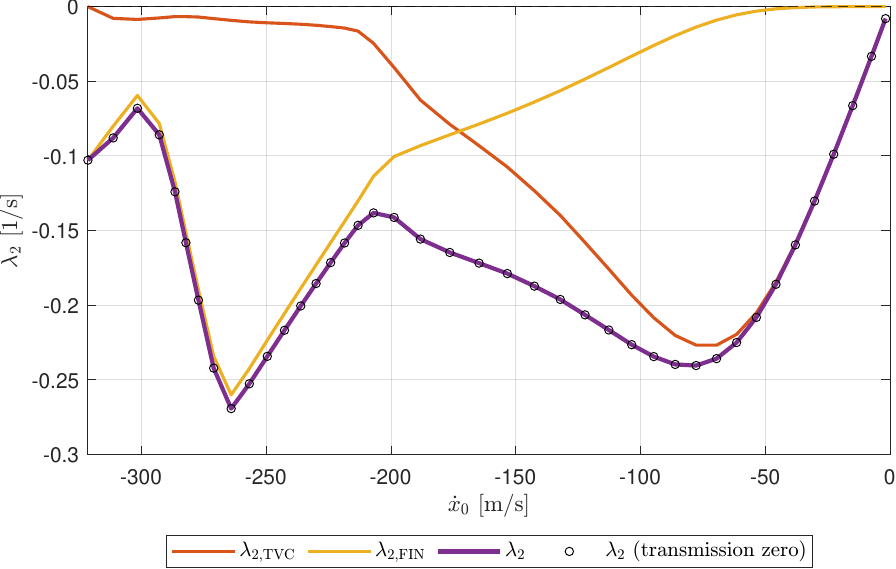}
    \caption{Internal-dynamics eigenvalue $\lambda_2$ and its TVC and fin contributions
    (Eq.~\eqref{eq:lambda_2_decomposition}), together with the non-zero transmission zero of
    the linear model.}
    \label{fig:internal_dynamics_eigenvalue}
\end{figure}

\par Since the zero dynamics of a linear(ized) system are governed by its transmission zeros \cite{isidori1995nonlinear}, $\lambda_1$ and $\lambda_2$ correspond to the transmission zeros of the transfer function from $\ddot{\theta}_c$ to $\theta$, obtained from Eq.~\eqref{linEOM} with the single-loop allocation: the zero at the origin corresponds to $\lambda_1$, and the non-zero zero to $\lambda_2$. Figure~\ref{fig:internal_dynamics_eigenvalue} shows this non-zero transmission zero along the nominal trajectory, which coincides to machine precision with Eq.~\eqref{eq:lambda_2_decomposition}. Both the TVC and fin contributions, and hence $\lambda_2$, remain negative throughout the flight envelope. Stability of the internal dynamics is, however, a necessary but not sufficient condition for using the single-loop configuration throughout the flight envelope, since its drift channel remains subject to the phase transition of Section~\ref{Alloc_PhaseChange}.

\section{Controller Assessment}
\label{sec:Comparison}
This section is divided into four parts. First, Section~\ref{sec:pd_controller} presents the baseline controller used for comparison, based on the architecture proposed in \cite{guadagnini2025end}. Section~\ref{Stability Analysis} then compares the INDI controller with the baseline controller in terms of stability. Section~\ref{sec:mcCampaign} describes the Monte Carlo campaign and presents the corresponding results. Finally, Section~\ref{discussion} provides an overall comparison of the two controllers, highlighting their respective advantages and disadvantages.

\subsection{Baseline Controller}

\label{sec:pd_controller}

\begin{figure}[t]
    \centering
    \includegraphics[scale=0.14]{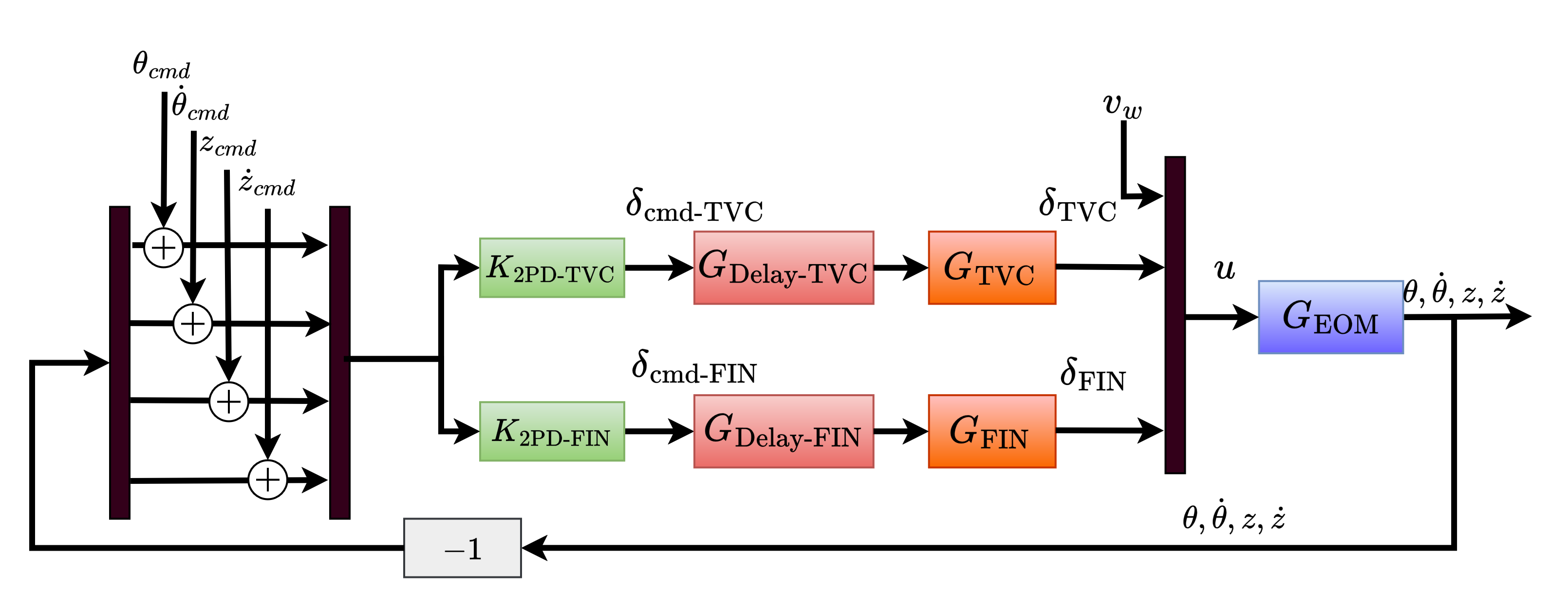}
    \caption{Block-diagram of the baseline control law.}
    \label{RetaltArch}
\end{figure}

Based on the work of \cite{guadagnini2025end}, the baseline controller architecture consists of four proportional-derivative (PD) controllers: two acting on the pitch angle and drift to generate the TVC command ($K_{2\text{PD-TVC}}$), and two acting on the same variables to generate the fin commands ($K_{2\text{PD-FIN}}$). The overall architecture is illustrated in Fig.~\ref{RetaltArch}. As with the previously presented INDI controller, the controller gains are gain-scheduled as a function of airspeed and kept constant outside the scheduling range, using the gains from the nearest design point.

\par The synthesis flight points are selected with the objective of achieving minimum loop-at-a-time input disk margins of $6\,\mathrm{dB}$ and $30^\circ$ throughout the flight envelope. This results in the need for six design flight points\footnote{At $\dot{x}_0=-112.7$\,m/s, the TVC input margin requirement could not be satisfied without degrading the terminal performance relative to the INDI controller, and is therefore marginally violated. Similarly, the fin input margin falls marginally below $6\,\mathrm{dB}$ between design points; given the small deviation, no additional flight point was introduced.}. The weighting functions used for controller synthesis are reported in Table~\ref{Hinftuning4PD}. Similarly to the INDI controller, at the first flight point, the margin requirements are enforced over a set of models generated from the corner cases of all LTI uncertainties, as this was found to be necessary to guarantee stability across all parameter combinations.

\par The control objectives, including their partition into hard and soft goals, are identical to those used for tuning the INDI-based controller (see Table~\ref{HinftuningINDI}). At coincident flight points, the tuning weights were set to the same values as with the INDI law, and it was ensured that the baseline controller satisfies all hard requirements at these flight points. At the remaining flight points, the weights were first obtained by interpolating between those of neighboring flight points and were then retuned to improve performance.
\par Finally, although the control architecture is identical to that proposed
in \cite{guadagnini2025end} for powered landing, the number of gain-scheduling
flight points has been significantly reduced, from more than \textit{twenty}
in the original implementation to only \textit{six} in the present work. The
proposed design therefore simplifies the original approach while retaining
the same underlying control architecture.
\begin{table}[b!]
\centering
\caption{$\mathcal{H}_\infty$ Closed Loop Shaping constraints used for baseline controller design.}
\begin{tabular}{l|cccccc|}
\cline{2-7}
\multicolumn{1}{c|}{} &
\multicolumn{6}{c|}{Flight Point (Airspeed in $\mathrm{m/s}$)} \\ \hline

\multicolumn{1}{|c|}{Constraints} &
\multicolumn{1}{c|}{$\dot{x}_0 = -249.6$} &
\multicolumn{1}{c|}{$\dot{x}_0 = -176.5$} &
\multicolumn{1}{c|}{$\dot{x}_0 = -112.7$} &
\multicolumn{1}{c|}{$\dot{x}_0 = -61.5$} &
\multicolumn{1}{c|}{$\dot{x}_0 = -37.9$} &
\multicolumn{1}{c|}{$\dot{x}_0 = -7.4$} \\ \hline

\multicolumn{1}{|l|}{LAT DB Margins} &
\multicolumn{1}{c|}{$2.5\,\mathrm{dB},\,12.5^\circ$} &
\multicolumn{1}{c|}{$6\,\mathrm{dB},\,30^\circ$} &
\multicolumn{1}{c|}{$6\,\mathrm{dB},\,30^\circ$} &
\multicolumn{1}{c|}{$6\,\mathrm{dB},\,30^\circ$} &
\multicolumn{1}{c|}{$6\,\mathrm{dB},\,30^\circ$} &
$6\,\mathrm{dB},\,30^\circ$ \\ \hline

\multicolumn{1}{|l|}{$\|T_{\theta_\mathrm{cmd} \rightarrow \delta_\text{TVC}}\|<W_1$} &
\multicolumn{1}{c|}{$0.17$} &
\multicolumn{1}{c|}{$2$} &
\multicolumn{1}{c|}{$2$} &
\multicolumn{1}{c|}{$2$} &
\multicolumn{1}{c|}{$2$} &
\multicolumn{1}{c|}{$2$} \\ \hline

\multicolumn{1}{|l|}{$\|T_{\theta_\mathrm{cmd} \rightarrow \delta_\text{FIN}}\|<W_2$} &
\multicolumn{1}{c|}{$4$} &
\multicolumn{1}{c|}{$4$} &
\multicolumn{1}{c|}{$4$} &
\multicolumn{1}{c|}{$4$} &
\multicolumn{1}{c|}{$4$} &
\multicolumn{1}{c|}{$4$} \\ \hline

\multicolumn{1}{|l|}{$\|T_{z_\mathrm{cmd} \rightarrow \delta_\text{TVC}}\|<W_3$} &
\multicolumn{1}{c|}{$0.17$} &
\multicolumn{1}{c|}{$2$} &
\multicolumn{1}{c|}{$2$} &
\multicolumn{1}{c|}{$2$} &
\multicolumn{1}{c|}{$2$} &
\multicolumn{1}{c|}{$2$} \\ \hline

\multicolumn{1}{|l|}{$\|T_{z_\mathrm{cmd} \rightarrow \delta_\text{FIN}}\|<W_4$} &
\multicolumn{1}{c|}{$4$} &
\multicolumn{1}{c|}{$4$} &
\multicolumn{1}{c|}{$4$} &
\multicolumn{1}{c|}{$4$} &
\multicolumn{1}{c|}{$4$} &
\multicolumn{1}{c|}{$4$} \\ \hline

\multicolumn{1}{|l|}{$\|T_{z_{\mathrm{cmd}} \rightarrow z}\|_\infty < W_5$} &
\multicolumn{1}{c|}{$\frac{1.90s}{s + 0.405}$} &
\multicolumn{1}{c|}{$\frac{1.3s}{s + 0.405}$} &
\multicolumn{1}{c|}{$\frac{1.08s}{s + 0.405}$} &
\multicolumn{1}{c|}{$\frac{1.185s}{s + 0.405}$} &
\multicolumn{1}{c|}{$\frac{1.32s}{s + 0.405}$} &
\multicolumn{1}{c|}{$\frac{4s}{s + 0.405}$}  \\ \hline

\multicolumn{1}{|l|}{$\|T_{\theta_{\mathrm{cmd}} \rightarrow \theta}\|_\infty < W_6$} &
\multicolumn{1}{c|}{$\frac{1.90(2s+0.01)}{s + 0.1}$} &
\multicolumn{1}{c|}{$\frac{0.55(2s+0.01)}{s + 0.1}$} &
\multicolumn{1}{c|}{$\frac{0.468(2s+0.01)}{s + 0.1}$} &
\multicolumn{1}{c|}{$\frac{0.593(2s+0.01)}{s + 0.1}$} &
\multicolumn{1}{c|}{$\frac{0.74(2s+0.01)}{s + 0.1}$} &
\multicolumn{1}{c|}{$\frac{1.4(2s+0.01)}{s + 0.1}$}  \\ \hline

\multicolumn{1}{|l|}{$\|T_{v_w \rightarrow z}\|_\infty < W_7$} &
\multicolumn{1}{c|}{$13.16$} &
\multicolumn{1}{c|}{$9$} &
\multicolumn{1}{c|}{$15$} &
\multicolumn{1}{c|}{$22.5$} &
\multicolumn{1}{c|}{$25$} &
\multicolumn{1}{c|}{$100$} \\ \hline

\multicolumn{1}{|l|}{$\|T_{v_w \rightarrow \dot{z}}\|_\infty < W_8$} &
\multicolumn{1}{c|}{$5.12$} &
\multicolumn{1}{c|}{$3.5$} &
\multicolumn{1}{c|}{$4$} &
\multicolumn{1}{c|}{$4.75$} &
\multicolumn{1}{c|}{$5$} &
\multicolumn{1}{c|}{$100$} \\ \hline

\end{tabular}
\label{Hinftuning4PD}
\end{table}

\subsection{Stability Analysis}
\label{Stability Analysis}

\begin{figure}[t!]
    \centering
    \begin{minipage}{0.49\linewidth}
        \centering
        \includegraphics[width=\linewidth]{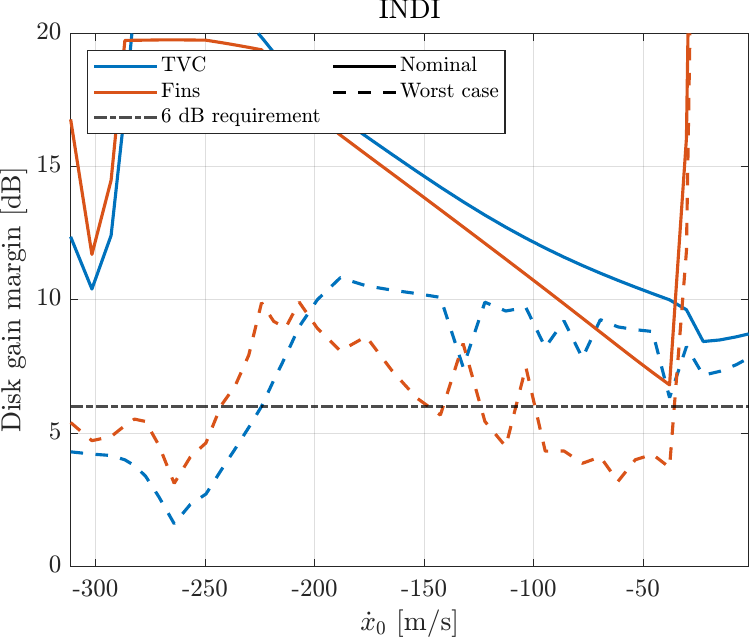}
    \end{minipage}
    \hfill
    \begin{minipage}{0.49\linewidth}
        \centering
        \includegraphics[width=\linewidth]{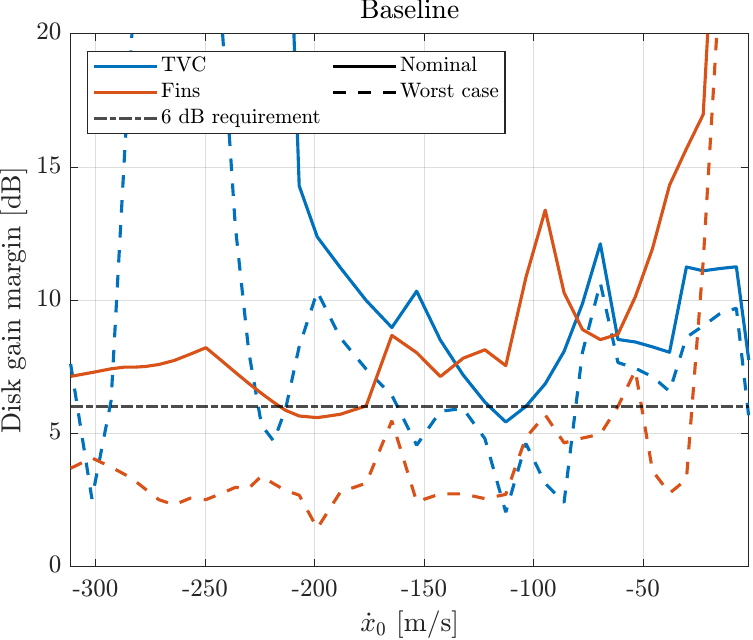}
    \end{minipage}
    \caption{Nominal and worst-case input disk-based stability margins of the INDI and baseline controllers.}
    \label{fig:INDIStabilityMargins}
\end{figure}

\begin{figure}[b!]
    \centering
    \begin{minipage}{0.49\linewidth}
        \centering
        \includegraphics[width=\linewidth]{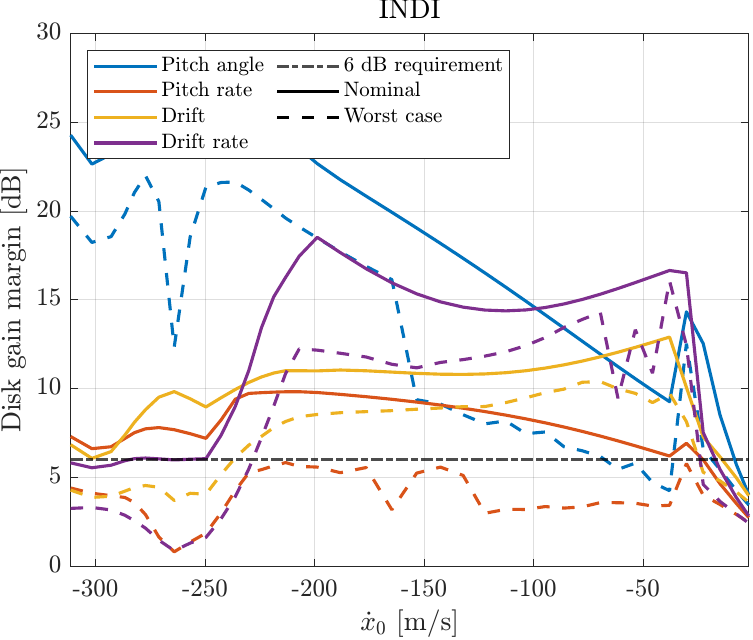}
    \end{minipage}
    \hfill
    \begin{minipage}{0.49\linewidth}
        \centering
        \includegraphics[width=\linewidth]{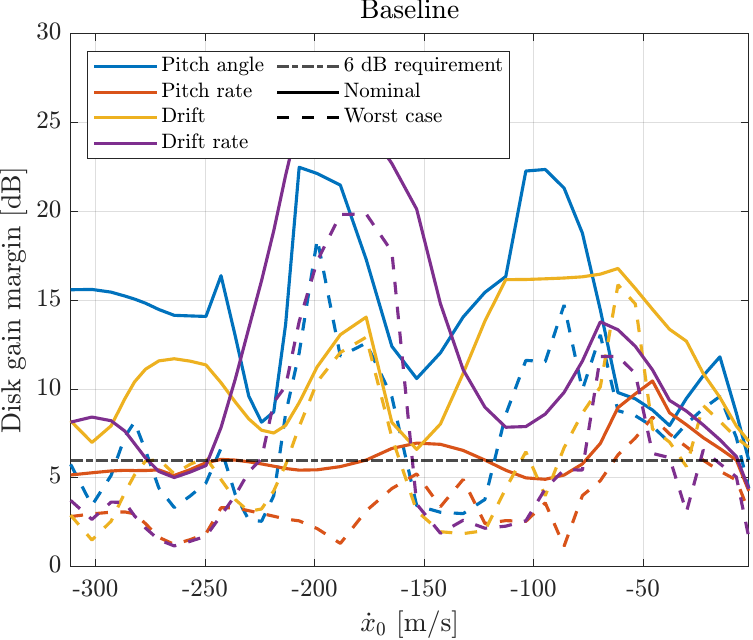}
    \end{minipage}
    \caption{Nominal and worst-case output disk-based stability margins of the INDI and baseline controllers.}
    \label{fig:BaselineStabilityMargins}
\end{figure}

In order to assess the stability of both controllers, input and output \textit{S-T} disk-based loop-at-a-time margins were evaluated under both nominal and dispersed conditions. For the dispersed cases, the worst-case disk-based margins were considered. The plant uncertainty used in this analysis is representative of the uncertainties considered in the Monte Carlo campaign. Figures~\ref{fig:INDIStabilityMargins} and \ref{fig:BaselineStabilityMargins} show the input and output margins, respectively, for the INDI and baseline controllers.

\par The INDI controller satisfies all input stability-margin requirements. The nominal margins remain well above the $6\,\mathrm{dB}$ requirement throughout the flight envelope, while the worst-case margins reach a minimum of approximately $2\,\mathrm{dB}$ but generally remain above $4\,\mathrm{dB}$, exceeding $6\,\mathrm{dB}$ in some regions. For the output margins, the nominal values remain above the $6\,\mathrm{dB}$ threshold over most of the flight envelope, with the exception of the drift rate channel at the very beginning of the envelope, where it remains marginally below this value, and of the very end of the envelope. Under worst-case conditions, the smallest output margins occur early in the flight envelope, where they drop to approximately $1\,\mathrm{dB}$. Thereafter, the worst-case margin for the pitch-rate output remains above $3\,\mathrm{dB}$, while the margins for the other outputs are considerably higher, except near the end of the flight envelope.
For the baseline controller, the nominal input stability-margin requirements are satisfied over most of the flight envelope, with one region in which the fin-related margin falls below the $6\,\mathrm{dB}$ requirement and another region in which the TVC-related margin falls below the same threshold. Nevertheless, the nominal input margins are generally significantly smaller than those of the INDI controller. The nominal output margins generally remain above the $6\,\mathrm{dB}$ threshold, with the exception of the pitch-rate channel, which falls below this value at several points. Near the end of the flight envelope, all four output margins fall below $6\,\mathrm{dB}$.

The worst-case margins are considerably lower than those obtained with the INDI controller. For the input margins, the lowest worst-case margin for the baseline controller remains close to $2\,\mathrm{dB}$ over most of the flight envelope, exceeding $3\,\mathrm{dB}$ only in three small regions: one in the middle of the flight envelope and two near the end. For the output margins, the minimum stability margin value occurs at approximately $-270$ to $-260\,\mathrm{m/s}$, similar to the INDI controller. However, after this minimum, the worst-case INDI margins remain, in general, above $3\,\mathrm{dB}$ for the pitch-rate channel and above $5\,\mathrm{dB}$ for the other output channels, whereas the corresponding worst-case output margins for the baseline controller consistently remain below $3\,\mathrm{dB}$ across all four output channels. Finally, the baseline controller's margins show substantially greater variation compared to those of the INDI controller. The smoother evolution observed for the INDI controller indicates a more consistent stability behavior across the flight envelope, whereas the larger variations in the baseline case reflect the stronger changes in closed loop behavior between flight points.

\subsection{Monte Carlo Campaign}
\label{sec:mcCampaign}
First, in Section~\ref{DefinitionMC}, the simulation environment is defined,
together with the modeling assumptions, uncertainty characterization, and
terminal requirements at landing. Subsequently, Section~\ref{MC_res} presents the Monte Carlo results for both controllers.
\subsubsection{Campaign Definition}
\label{DefinitionMC}

\begin{table}[b!]
\centering
\caption{Monte Carlo dispersion parameters.}
\label{tab:MCdisp}
\renewcommand{\arraystretch}{1.15}
\begin{tabular}{lcc}
\hline
\textbf{Parameter} & \textbf{Dispersion} & \textbf{Distribution} \\
\hline

Initial propellant mass, $m_0$ 
& $\pm 2.5\%$ 
& Normal (1-sigma) \\

Initial position, $\mathbf{r}_0$
& $\|\Delta\mathbf{r}_0\| = 65~\mathrm{m}$
& Normal (3-sigma) \\

Initial velocity, $\mathbf{v}_0$
& $\|\Delta\mathbf{v}_0\| = 10~\mathrm{m/s}$
& Normal (3-sigma)\\

Meridional wind 
& $\pm 2.5\ \,\mathrm{m/s}$ 
& Uniform \\

Zonal wind 
& $\pm 4.5\ \,\mathrm{m/s}$ 
& Uniform \\

Axial aerodynamic coefficient, $C_A$ 
& $\pm 25\%$ 
& Uniform \\

Normal-force coefficient, $C_N$ 
& $\pm 25\%$ 
& Uniform \\

Pitch-moment coefficient, $C_M$ 
& $\pm 25\%$ 
& Uniform \\

Diagonal inertia terms 
& $\pm 3\%$ 
& Uniform \\

Off-diagonal inertia terms 
& $\pm 30\,\mathrm{kg\,m^2}$ 
& Uniform \\

Center-of-gravity position 
& $\pm\begin{bmatrix}1 & 0.1 & 0.1\end{bmatrix}^{\top}\,\mathrm{m}$ 
& Uniform \\

\hline
\end{tabular}
\end{table}

The performance and robustness of the two controllers are assessed through a Monte Carlo campaign conducted using the ESA-i4GNC simulation framework \cite{ESAi4gnc} and
the RETALT reusable launch vehicle parameters described in
\cite{guadagnini2025end}. The simulations employ the complete nonlinear
six-degree-of-freedom vehicle dynamics during the powered landing phase,
under the assumption of a non-rotating Earth.
\par Atmospheric disturbances are introduced through the meridional and zonal wind components, corresponding to the north-south and east-west directions, respectively. Uncertainty is added to the nominal wind profile according to the dispersions reported in Table~\ref{tab:MCdisp}, which are defined relative to the reference atmospheric profile.

\begin{figure}[t!]
    \centering
    \includegraphics[width=0.65\linewidth]{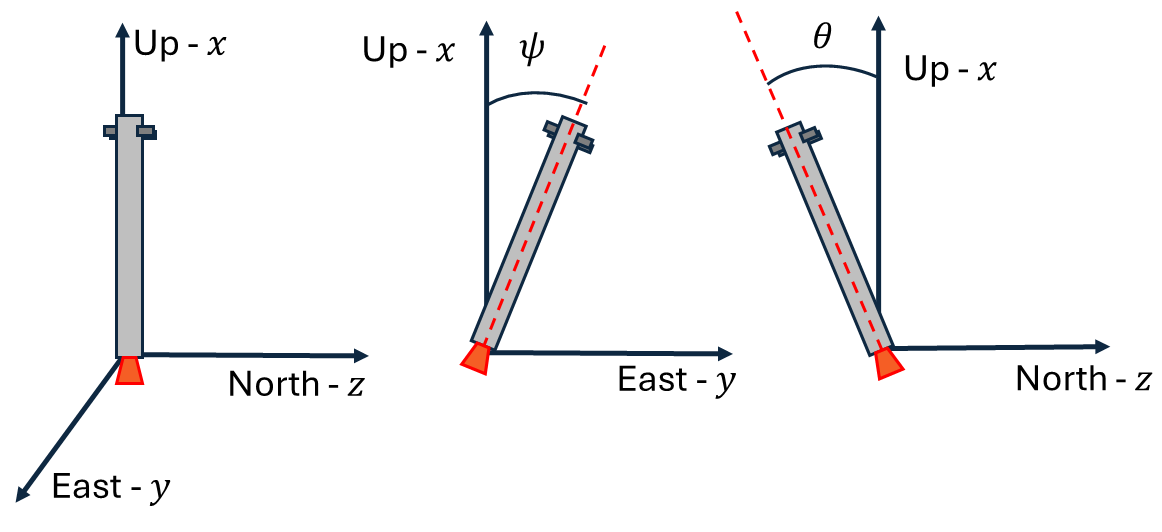}
    \caption{Axis definition and pitch/yaw angle conventions: three-dimensional view (left), $x$-$y$ plane projection (center), and $x$-$z$ plane projection (right).}
    \label{fig:AxisOrientation}
\end{figure}

\par The vehicle model accounts for the TVC and fin actuator dynamics, as
well as sensor noise and bias. Flexible-body dynamics and propellant sloshing are neglected. These effects have yet to be addressed within the proposed INDI framework; however, they are expected to be handled in a similar manner as in NDI-based designs for flexible vehicles, namely through appropriate filtering in the forward path~\cite{gregory2005}.
\par The GNC subsystem operates at $50\,\mathrm{Hz}$. For each simulation,
the lateral position and attitude control is provided by one of the two
controllers considered in this work, while the longitudinal and roll
controllers are those described in \cite{guadagnini2025end}. Guidance is
provided by the Successive Convexification (SCvx) approach described in
\cite{CEAS-GNC-2026-047}, with the reference trajectory generated once at
the beginning of each simulation and subsequently kept fixed. Navigation is
represented by a performance model, in which the relevant quantities are
estimated with sufficient accuracy while accounting for the sensor noise
and bias described above.

\par For each controller, a total of $500$ Monte Carlo runs are performed, consistent with previous studies using the same
vehicle and simulation framework \cite{guadagnini2025end,Diz_2026_esagnc}.
The campaign considers dispersions in the initial conditions, atmospheric
disturbances, aerodynamic coefficients, and MCI properties. The dispersed
parameters and their corresponding uncertainty levels are listed in
Table~\ref{tab:MCdisp}, with all uncertain quantities assumed to be
statistically independent.

\par Finally, a simulation is considered successful if the terminal
requirements at touchdown are satisfied. These requirements are defined in
terms of the final position, velocity, attitude, and attitude-rate errors
with respect to the reference landing condition. The reference axes and the pitch and yaw angles $\theta$ and $\psi$ used to define these requirements, are shown in Fig.~\ref{fig:AxisOrientation}. Note that, unlike in Section~\ref{sec:linear_model}, where $\theta$ is measured from the local horizontal, the angles used for the terminal requirements are measured from the local vertical, such that the nominal touchdown attitude corresponds to $\theta=\psi=0$. The lateral terminal requirements are summarized in Table~\ref{tab:terminal_constraints}
\footnote{The adopted values are based on those used in
\cite{guadagnini2025end}, except for the velocity-error requirement, which is
considerably more stringent.}.

\begin{table}[b!]
\centering
\caption{Terminal requirements for the Monte Carlo campaign.}
\label{tab:terminal_constraints}
\renewcommand{\arraystretch}{1.15}
\begin{tabular}{lcc}
\hline
\textbf{Quantity} & \textbf{Components} & \textbf{Requirement} \\
\hline

Position error
& $(y,z)$
& $\|(y,z)\|_2 < 10\,\mathrm{m}$ \\

Velocity error
& $(\dot y,\dot z)$
& $\|(\dot y,\dot z)\|_2 < 1\,\mathrm{m/s}$ \\

Attitude error 
& ($\theta$, $\psi$) 
& $\|(\theta,\psi)\|_2 < 5^\circ$ \\

Attitude rate error 
& $(\dot\theta,\dot\psi)$ 
& $\|(\dot{\theta},\dot{\psi})\|_2 < 1^\circ/\mathrm{s}$ \\
\hline
\end{tabular}
\end{table}

\begin{figure}[t!]
        \centering
        \includegraphics[width=0.8\linewidth]{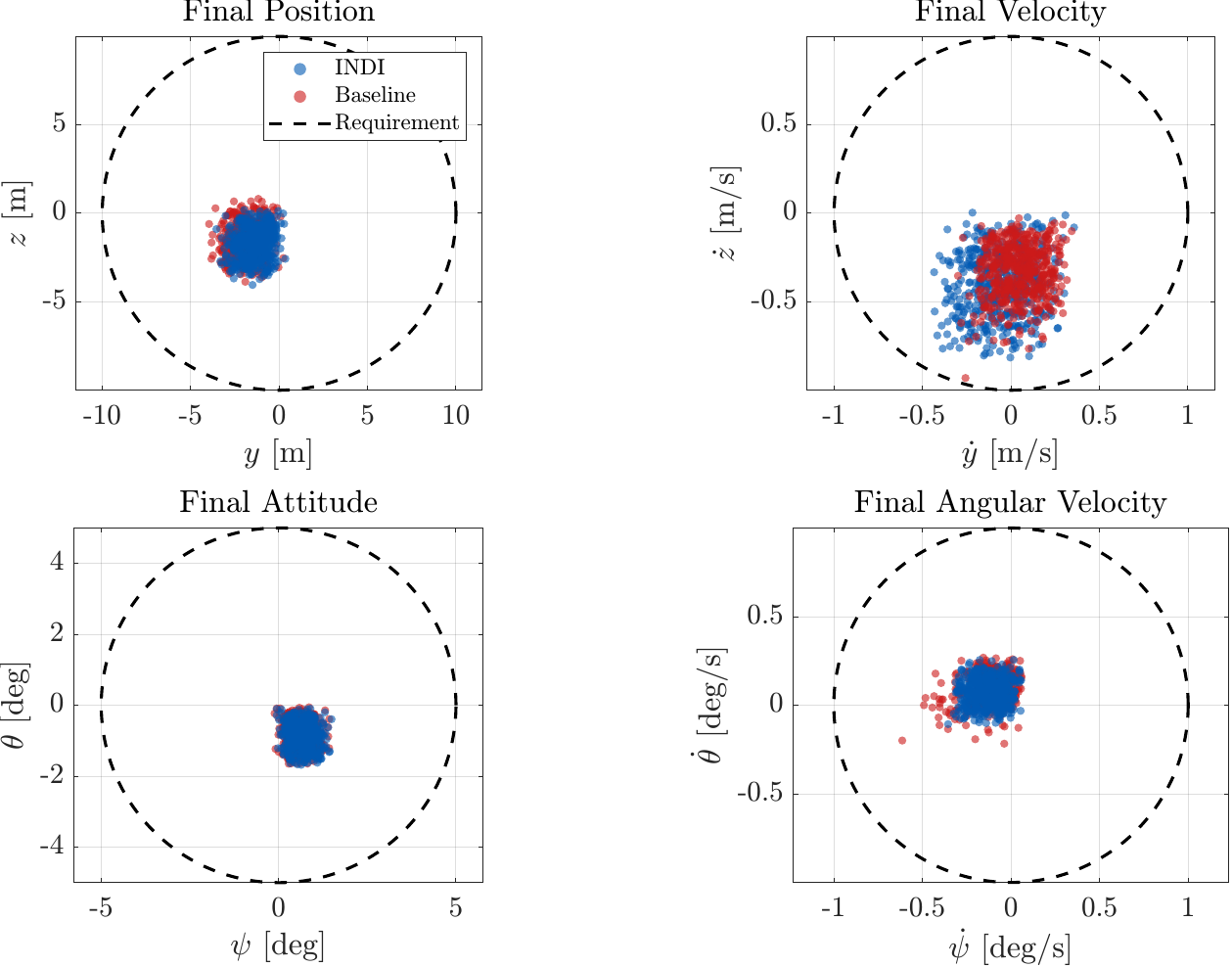}
        \caption{Terminal results at landing for the INDI and baseline controllers.}
        \label{fig:MC_results_INDI}
\end{figure}

\subsubsection{Results}
\label{MC_res}

This subsection presents the Monte Carlo results obtained with the INDI and baseline controllers under the dispersed conditions defined previously. The terminal results for the INDI and the baseline controllers are superimposed and shown in Fig.~\ref{fig:MC_results_INDI}.

\par As shown in Fig.~\ref{fig:MC_results_INDI}, all terminal requirements are satisfied in all $500$ simulations with the INDI controller. The final positions are clustered within a radius of approximately $2\,\mathrm{m}$, slightly offset from the target due to the preferential wind direction. The final velocity remains within the $1\,\mathrm{m/s}$ requirement, while the final attitude remains within $2^\circ$. The final angular velocity is within $0.35^\circ/\mathrm{s}$ in all but one simulation. The baseline controller also satisfies all terminal requirements in all $500$ simulations, with the final positions showing a similar dispersion and offset. The final velocity and attitude remain within their respective $1\,\mathrm{m/s}$ and $5^\circ$ requirements, while the final angular velocity remains within $0.5^\circ/\mathrm{s}$ in all but one simulation run.
 \begin{table}[b!]
    \centering
    \caption{Standard deviation of lateral terminal conditions over the 500-case Monte Carlo campaign.}
    \label{tab:MC_terminal_STD}
    \begin{tabular}{lcc}
        \hline
        Terminal Quantity & INDI STD & Baseline STD \\
        \hline
        Final Position [m] & 0.8614 & 0.8895 \\
        Final Velocity [m/s] & 0.1703 & 0.1427 \\
        Final Attitude [deg] & 0.3566 & 0.3583 \\
        Angular Velocity [deg/s] & 0.0761 & 0.0874 \\
        \hline
    \end{tabular}
\end{table}
The two controllers exhibit only minor differences in the terminal results. For the final position, the INDI controller achieves slightly greater accuracy and lower variation across the simulations. Conversely, the baseline controller achieves better final velocity performance; this difference is consistent with the INDI controller maintaining a stronger emphasis on position control through the end of the maneuver, resulting in slightly higher residual velocity. Nonetheless, it should be noted that the baseline controller comes extremely close to violating the lateral terminal velocity requirements in one of the simulations, whereas the INDI controller retains a larger margin to the requirement. For the attitude, both controllers exhibit nearly identical results, while for the angular velocity, the INDI controller produces smaller values and less variation across the simulations. The standard deviation of the different results can be seen in
Table~\ref{tab:MC_terminal_STD}. Overall, both controllers successfully satisfy all terminal requirements in all $500$ Monte Carlo shots, demonstrating equivalent robust performance under the considered dispersions.

\begin{figure}[t!]
    \centering
    \includegraphics[width=\linewidth]{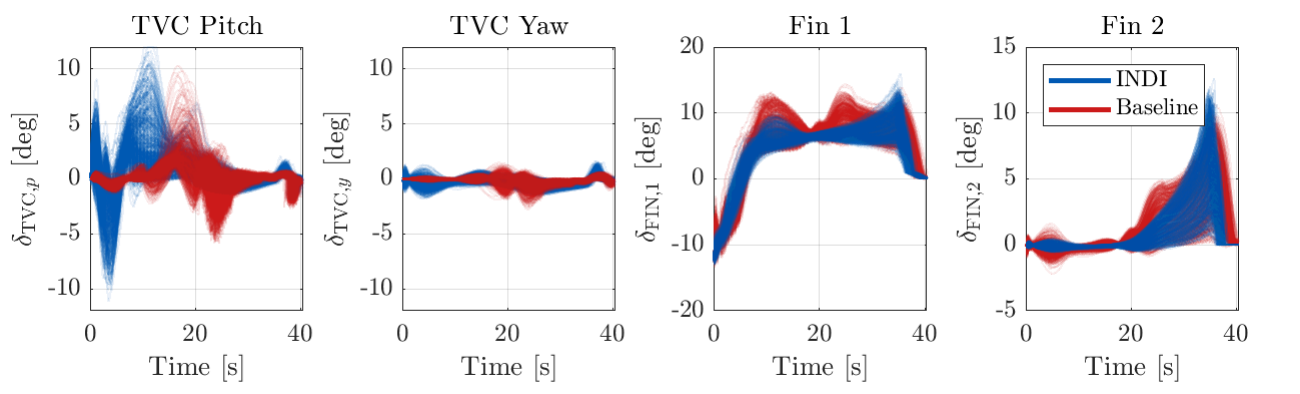}
    \caption{TVC and fin deflections over the Monte Carlo simulations for the INDI and baseline controllers.}
    \label{fig:actMC_results_INDI}
\end{figure}

\begin{figure}[b!]
        \centering
        \includegraphics[width=0.8\linewidth]{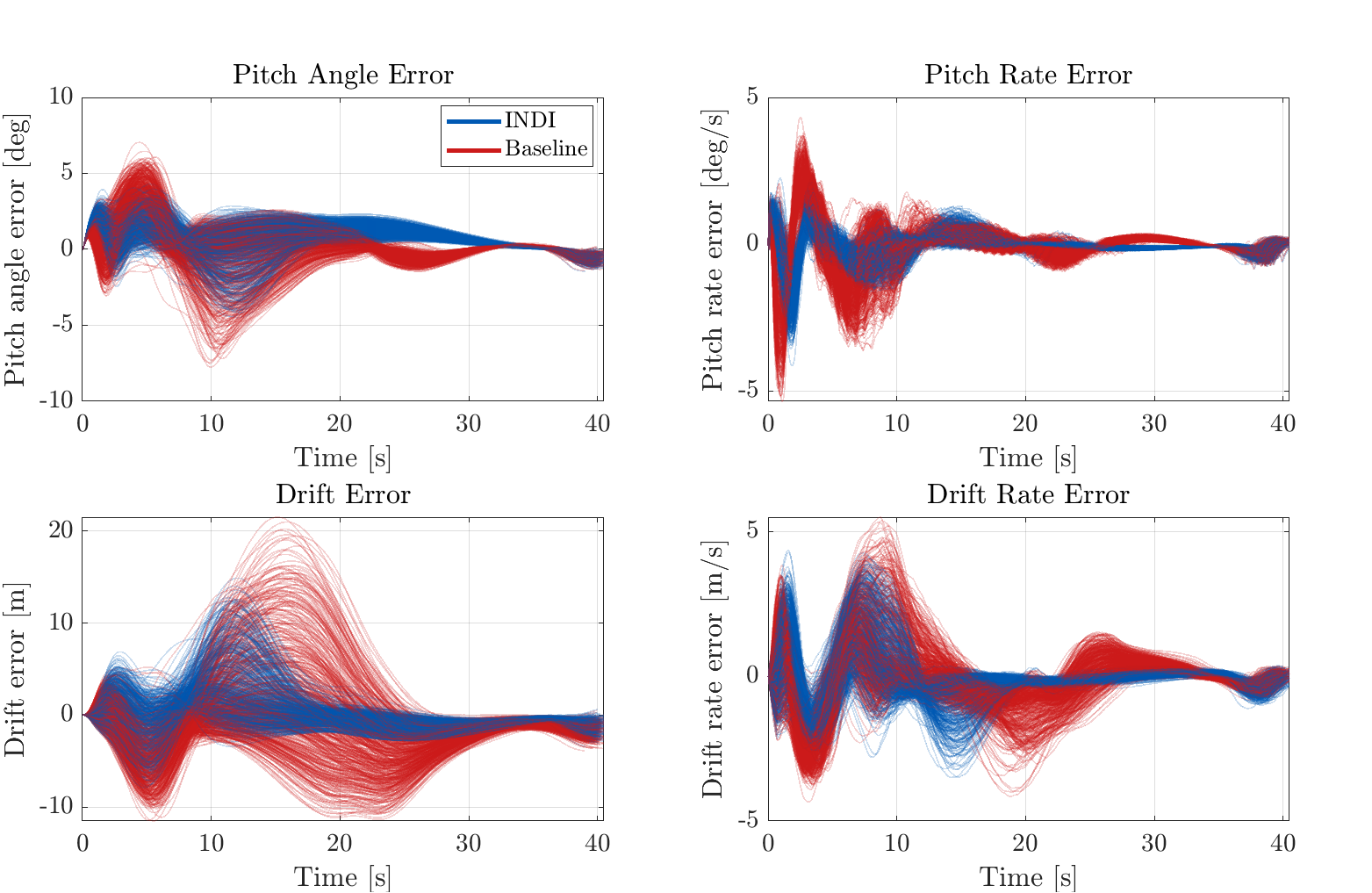}
        \caption{Pitch plane error evolution for the INDI and baseline controllers.}
        \label{fig:MC_results_baseline}
\end{figure}

The corresponding actuator deflections are shown in Fig.~\ref{fig:actMC_results_INDI}. Fin usage is broadly comparable between the two architectures, although the baseline controller exhibits a larger shaded area, indicating greater fin usage overall. In contrast, the TVC demands differ more significantly, particularly during the early portion of the flight envelope, where the INDI controller commands substantially larger TVC deflections. This increased actuation can be attributed to the improved trajectory tracking achieved by the INDI controller. Figure~\ref{fig:MC_results_baseline} illustrates the evolution of the pitch plane errors across all simulations, including the pitch angle error, pitch rate error, drift error, and drift rate error. For all four quantities, the INDI controller not only exhibits smaller peak errors and lower dispersion across the simulations, but, most importantly, achieves significantly faster convergence to errors near zero.

Overall, the two controllers exhibit comparable terminal performance but substantially different tracking capabilities. The INDI controller follows the reference trajectory more closely and achieves this with considerably greater consistency across the dispersed simulations. This improved tracking performance suggests greater potential for accurately following more demanding reference trajectories, particularly those involving larger or faster variations in the commanded states.

\subsection{Discussion}
\label{discussion}

The comparison between the INDI-based and baseline controllers leads to several observations. In terms of robust performance, the two architectures achieve comparable terminal conditions and can therefore be considered broadly equivalent with respect to the final state. However, clear differences are observed in the evolution of the error variables throughout the trajectory. The INDI-based controller exhibits substantially lower error variation, lower peak errors, and significantly faster convergence toward zero, resulting in improved trajectory tracking. In terms of stability, as discussed in Section~\ref{Stability Analysis}, the INDI-based controller also provides significantly larger nominal and worst-case stability margins. A further distinction is observed in the number of gain-scheduling flight points required to achieve sufficient stability and performance. The INDI-based controller requires only three flight points, compared with six for the baseline controller. \smallskip
\par The two architectures also differ in controller complexity. The baseline controller contains eight tunable gains, whereas the INDI-based controller requires four first-order filters and six gains. One of the filters, used to estimate the linear acceleration, could be removed since this measurement is directly available; it was retained only to ensure a fair comparison with the baseline controller. Although the first-order filters introduce additional controller complexity, they are tuned only once and subsequently kept fixed throughout the flight envelope. Their contribution to the overall tuning complexity is therefore limited.
\smallskip
\par The largest difference, however, arises in the tuning process. The
INDI-based architecture was significantly simpler to tune and more permissive
with respect to the choice of tuning weights, as even non-optimal weight
selections resulted in satisfactory controller performance. This is due to the
fact that the inner inversion loop provides a favorable baseline loop gain
\cite{Diz_2026_eurognc}. Although the inner-loop alone does not necessarily
stabilize the system, it brings the loop gain close to the desired shape, such
that the outer-loop only needs to provide the additional gain required for
stabilization and to fine-tune performance. The baseline architecture does not inherently provide this favorable baseline loop gain and consequently offers greater freedom in controller tuning. This greater freedom also makes the tuning process more sensitive to the selection of tuning weights and increases the effort required to achieve the desired performance and robustness characteristics across the flight envelope.
\smallskip
\par The INDI architecture also offers advantages in terms of missionization and extension to other flight stages. The reduced number of flight points simplifies the design, while the feedback linearization can reduce the retuning required when adapting the controller to a different mission or vehicle. This normalization and linearization can also be leveraged to operate linear online guidance strategies, as demonstrated in \cite{Diz2026PostCapture}. Furthermore, the architecture can be readily adapted to changes in the available control effectors. For example, during an aerodynamic descent phase where only the fins are available, the single-loop configuration explored in this work can be leveraged. In contrast, the baseline architecture provides no inherent mechanism for such adaptation and would generally require a new controller design or substantial retuning for different missions, vehicles, flight stages or actuator configurations.

\section{Conclusion}
\label{sec:conclusions}
This work first investigated the lateral dynamics of reusable launch vehicles during powered descent and identified a fundamental open-loop limitation that arises during its final phase. The analysis established an analytical condition governing the occurrence of an abrupt $180^\circ$ phase transition in the transfer functions from the control inputs to drift and drift rate. The underlying mechanism was given a physical interpretation: as the vehicle decelerates during the final phase of descent, the lateral drift dynamics transition from an aerodynamically dominated regime to one in which the contribution associated with the gravity-dependent coupling between attitude and drift becomes dominant, resulting in a reversal of the effective input-output sign. Since the transition depends on vehicle mass, aerodynamic properties, and geometry, its location can be influenced through early vehicle and mission design choices, allowing it to be shifted away from critical flight phases. Importantly, the analysis shows that the identified limitation originates from the underlying vehicle dynamics rather than from a particular controller implementation. Consequently, it must be accounted for by any control architecture regulating lateral drift on an RLV that undergoes a similar transition from an aerodynamically dominated regime to one in which the gravity-dependent attitude-drift coupling becomes dominant during powered descent.
\par To address the fundamental limitation, an INDI-based control law was
developed. The proposed architecture was shown analytically to address the
limitation inherently through the decoupling of attitude and position.
Furthermore, the internal dynamics introduced by the feedback linearization were derived and proven to be stable throughout the considered flight envelope. Together, these results provide a sound theoretical foundation for
the proposed control law. Its performance was then assessed through a Monte
Carlo campaign and compared against a baseline controller consisting of four
gain-scheduled proportional-derivative controllers. Both controllers achieved
a $100\%$ success rate over the considered Monte Carlo campaign, demonstrating
comparable robust terminal performance. However, during the transient response, the INDI-based controller exhibited substantially lower error
variation and lower peak errors, while converging to near-zero errors
significantly faster. The faster error convergence suggests that the INDI-based architecture is better suited to tracking more aggressive trajectories. The INDI controller also provided superior nominal and
worst-case stability margins, required fewer gain-scheduling flight points,
and substantially reduced the associated tuning effort. The architecture further offers advantages for missionization and extension to
other flight stages, as the INDI formulation facilitates the
adaptation of the controller to different missions, vehicles, and available
control effectors.

\section*{Funding Sources}

This work was funded by the European Space Agency through Contract No. 4000147243/25/NL/MGu/ov.

\section*{Acknowledgments}
The first author would like to thank the members of the INDRA Space GNC team for their valuable discussions and support, in particular Pietro Ghignoni, Afonso Botelho, and João Belfo, whose insight and continued support were especially important throughout this work. The first author also thanks Luis Basabe and Guilherme João Videira for the helpful exchanges and their assistance with the simulation environment. A generative AI tool was used for proofreading the manuscript and for minor modifications of the authors' original diagram in Fig.~\ref{fig:LVPitchDiagram}.

\bibliography{sample}

\end{document}